\documentstyle[prb,twocolumn,aps,epsf,psfig]{revtex}
\begin{document}
\twocolumn[\hsize\textwidth\columnwidth\hsize\csname
@twocolumnfalse\endcsname
\draft
\title{Phase separation frustrated by the long range Coulomb  
interaction I: Theory} 
\author{J. Lorenzana,$^{1,2}$ C. Castellani,$^1$ and C. Di Castro$^1$}
\address{
$^1$ Dipartimento di Fisica, Universit\`a di Roma "La Sapienza'' and  
Istituto Nazionale di Fisica della Materia, 
Unit\`a di Roma I,\\
Piazzale A. Moro 2, I-00185 Roma, Italy.}
\address{
$^2$ Consejo Nacional de 
Investigaciones Cient\'ificas y Tecnicas, Centro At\'omico Bariloche,
8400 S. C. de Bariloche, Argentina }
\date{\today}
\maketitle
\begin{abstract}
We analyze the combined effect of the long range Coulomb (LRC) interaction 
and of surface energy on  first order density-driven phase transitions 
between two phases  in the presence of a  compensating rigid background.   
In the coexistence region we study  mixed states formed by regions of
one phase surrounded by the other  in the case in which the scale of the 
inhomogeneities is much larger than the interparticle distance.
Two geometries are studied in detail: spherical drops of one phase
into the other and a layered structure of one phase alternating with
the other. In the latter case we find the optimum density profile in
an approximation in which the free energy  is a functional of the
local density (LDA). It is
shown that an approximation in which the density is assumed to be
uniform  (UDA) within each phase region gives results very similar
 to those of the  more involved LDA approach.  Within the
UDA  we derive
the general equations for the chemical potential and the pressures of 
each phase  which generalize the Maxwell
construction to this situation. The equations are valid for a rather
arbitrary geometry.  We find that the transition 
to the mixed state is quite abrupt i.e. inhomogeneities  of the first phase 
appear with a 
finite value of the  radius and of the phase volume fraction. The maximum size
of the inhomogeneities is found to be on the scale of a few electric
field screening lengths. Contrary to the 
ordinary Maxwell construction, the inverse specific volume of each phase 
depends here on the global density in the coexistence region and can decrease
as the global density increases. 
  The range of densities in which coexistence is observed shrinks as the 
 LRC  interaction increases until 
it reduces to a  singular point. We argue that close to this 
singular point the system undergoes a lattice instability  
as long as the inverse lattice compressibility 
is finite. 
\end{abstract}

\vskip2pc]

\narrowtext
\section{Introduction}

The complex phase-diagrams of hole doped cuprates and manganites have rekindled
the study of mixed states  in modeling these 
systems.\cite{mul92,sig93,mor99} 
Indeed strongly correlated systems with narrow bandwidth 
and short range interactions show a generic tendency
to phase separate  into  hole-rich and 
hole-poor  regions. When long range Coulomb (LRC) forces are taken into 
account this  instability with 
macroscopic separation is frustrated due to the electrostatic energy cost
and this can lead to charge inhomogeneous states of various nature,\cite{nag83,nag98,low94,cas95b}
where domains of various forms of one phase ($B$) are embedded in the other
phase ($A$).

In the inhomogeneous state the charge 
is  segregated locally over some characteristic distance 
but the overall density (averaged over much larger distances) 
is a fixed constant in order  to guarantee large scale 
neutrality and avoid the large Coulomb cost.  Such a segregation has
been  considered at a scale comparable to the interparticle distance
to explain the origin of 
striped states in cuprates.\cite{low94,cas95b}

In this work we will consider the opposite case in which the 
scale of the inhomogeneities is much larger than the  interparticle
distance. We consider in particular two different kinds of
inhomogeneities: spherical drops of one phase  into the other phase
and alternating layers of each phase.  
The first case has been pioneered by Nagaev and collaborators in the 
context of doped magnetic semiconductors in general and of 
manganites in particular.\cite{nag83,nag98}
Related ideas have been recently presented in Refs.~\onlinecite{mor99,kag00}.

We believe that for the general understanding of the large scale 
inhomogeneous state the specific mechanism producing phase separation (PS)
in the absence of LRC forces is rather unessential.
Of course  specific short range interactions  in each 
physical system will lead to different $A$ and $B$ phases (which  will also  
depend on the doping regions one considers) giving rise  to different 
physical situations. However, in the same spirit of the Maxwell
construction, one can perform a general analysis of the phenomena due to the tendency
towards PS in the presence of Coulomb interaction irrespectively of the 
microscopic mechanisms of PS itself.

We consider two charged phases $A$ and $B$ with a compensating rigid background
and  we study the  formation of inhomogeneous states in a
density-driven first-order phase 
transition  between $A$ and $B$. By definition $A$ and $B$ have different 
densities; one of the phases is undercompensated and the other is
overcompensated by the background. It follows that the inhomogeneities are
charged and they repel each other. Since the inhomogeneities   
are formed by many particles, quantum effects are negligible and they
crystallize.\cite{notequa} The drops arrange in a  Wigner
crystal  whereas the layers form a periodic structure. We restrict to 
three dimensional (3D) textures. A large number of small inhomogeneities 
minimize the Coulomb energy but they cost too much surface energy. 
The distance between the inhomogeneities
and their size are found by minimizing a free energy which takes
into account both these effects.\cite{nag83,nag98}

In ordinary PS  the Maxwell construction (MC) is  
invoked to find the range of density  $n_A^0<n<n_B^0$ in which a system 
prepared with the overall 
density $n$ separates in two regions with densities $n_A^0$ and $n_B^0$ 
respectively. We generalize here the MC and 
derive the equations that  should be satisfied
in the mesoscopically inhomogeneous coexistence  region by 
the chemical potential and the pressure of each phase 
(Sec.~\ref{sec:freee}). To this end we use an approximation in which 
the  density within each phase is assumed to be constant 
which we name uniform density approximation (UDA). 
We solve the equations for the drop geometry 
in the simple (but general enough case) in which the free energy of both the 
$A$ and $B$ phases can be approximated by a 
parabola (Sec.~\ref{sec:parabolic}). 

We define a coupling constant $\lambda$ given by the ratio between the
the energy cost due to surface energy plus the LRC interaction 
and the  energy gain in MC PS.  Only below a critical value $\lambda_c$ PS is
possible.  Above $\lambda_c$ the system is uniform $A$ ($B$) phase below 
(above)  a
critical density with a lattice instability close to the critical density.

The characteristic size of the inhomogeinites is shown to be of order
$\sqrt{\lambda} l_s$ with $l_s$ an electric field screening length.
Since $\lambda$  is bounded  by  $\lambda_c$ (of order 1) it follows that the 
inhomogeneities are of the order of or less than $l_s$.

For small volume fraction  the drop geometry is more stable than the layers
as expected from general
surface energy arguments. On the other hand the layered  geometry 
being  simpler serves as a  ground test for approximations. 
In order to validate the UDA we solve the layered geometry in the UDA  
and  in the more general case in which the
density profile  can spatially vary within each phase. In this last case the density 
profile is allowed to adjust  minimizing a 
free energy which is an approximate functional of the local 
density (LDA). Both the UDA and the LDA  are shown to give very
similar results for averaged quantities  (Sec.~\ref{sec:tf}).

To illustrate the generality of theses ideas we consider 
some applications in paper II of this series.

\section{Free Energy and coexistence Equations: The uniform density 
approximation}
\label{sec:freee}

We consider a density-driven first order phase transition in the presence 
of the LRC interaction and surface energy. We look for the formation of 
a mixed state by increasing the density from the uniform A phase. 
We use two different geometries for the mixed state. i) The drop geometry
consist of a Wigner crystal of drops of $B$ phase in the 
host phase $A$. ii) The layered geometry is made of a periodic array of
alternating layers of $A$ and $B$ phases.

For both geometries  the electronic density
within  each single phase region is taken as uniform (UDA) and in general it will result different from the compensating background density.
This is of course an approximation since both densities 
will  tend to adjust within each phase also to make the total 
electrochemical potential constant. 
The UDA  will be relaxed in Sec.~\ref{sec:lda}
for the layered geometry by minimizing a free energy functional
on a simple LDA.  We anticipate here
that both the UDA and the LDA give very similar results thus 
justifying our extensive use of the UDA here and in paper II.  

We start by computing the total free energy.  In the same spirit of
the MC we assume that the free energies of hypothetically homogeneous
bulk phases are known and given by $F_A$ and $F_B$.  We define the
mixing energy $E_m$ as the sum of the total surface energy and
electrostatic energy (computed below).  We work at a fixed total
volume $V$ and number of particles $N$.  At a given temperature the
total free energy is $F=F_B(V_B,N_B)+F_A(V_A,N_A)+E_m$.  We have to
minimize this respect to $V_B$ and $N_B$ subject to the conditions
$V_B+V_A=V$, $N_B+N_A=N$. The volume fraction of $B$ phase is
$x\equiv V_B/V$. We can work with the free energies per unit volume
$f\equiv F/V$, $e_m\equiv E_m/V$, $f_B\equiv F_B(V_B,N_B)/V_B$ and
work with the densities $n_B\equiv N_B/V_B$ etc. so the function to
minimize is:
\begin{equation}
\label{eq:fdx}
f=(1-x)f_A(n_A)+x f_B(n_B)  + e_m
\end{equation}
The constraint in the number of particles is written as 
$n=x n_B+(1-x) n_A$ and the constraint in the volume is  satisfied by 
putting $V_B/V=1-x$. It is convenient to define $\delta\equiv n_B-n_A$ and 
to use the constraint in the number of particles 
to eliminate $n_B$ and $n_A$ in favor of $n$ and $\delta$.

\begin{figure}[tbp]
\epsfverbosetrue
\epsfxsize=8cm
$$
\epsfbox{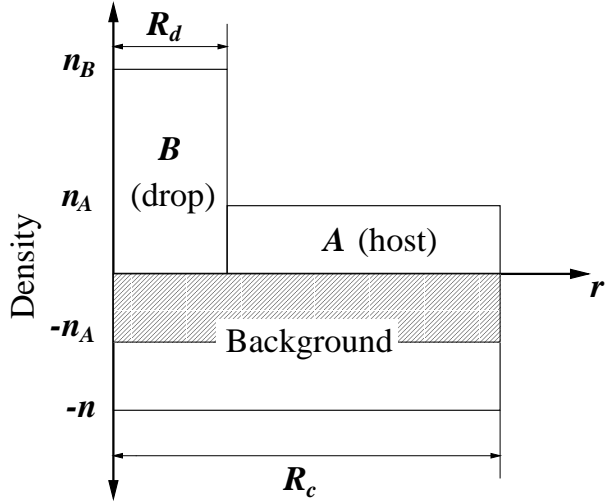}
$$
\caption{Schematic view of a cell density profile in the UDA with a 
drop (layer) of B phase in the host A. The 
origin is at the center of the cell. The full cell diameter (width)
is $2R_c$ for drops (layers).  
 The dashed region of the
background compensates the $A$ density and part of the $B$ density.  }
\label{fig:dropdr}
\end{figure}

In order to compute the mixing energy we first  consider the drop geometry. 
We  assume that the drops are spheres of radius $R_d$. This will be a good
approximation as long as $x$ is small and the crystal field
is also approximately spherical. This is true for 
fcc, bcc and hcp lattices.\cite{wig34,mah90} To compute the electrostatic
energy we use the Wigner-Seitz approximation.\cite{nag83,wig34,mah90}
We divide the system in slightly overlapping spherical cells each one with the 
volume $4\pi R_c^3/3=V/N_d$ where $N_d$ is the number of drops
and $R_c$ is the radius of the cell. 
Fig.~\ref{fig:dropdr} shows a schematic view of the cell density profile.

Next  we compute the electrostatic energy. The cells are globally neutral
(by construction) and only the charge inside the
cell contributes to the electric field in the cell. 

The charge density of phase $B$ is $n_B$ (actually $-e n_B$ but we drop
the charge of the particles $-e$ for simplicity). 
The dashed background  charge density in  Fig.~\ref{fig:dropdr} ($-n_A$) 
compensates the $A$ charge density $n_A$, and a 
slice of height $ n_A$ of the $B$ charge density. For the purpose of 
computing the electrostatic energy these charge densities can be 
eliminated and one is left with the density $n_B-n_A(=\delta)$ inside the
drop  and $ -(n-n_A)$ for the background. We will call the former ``drop
contribution'' and the latter the ``background contribution''. There is no
``host'' contribution due to the above cancellation.

Another assumption is that the charge is spread uniformly and that 
microscopic  discreetness effects can be neglected. One can see that 
corrections to the electrostatic energies due to discreetness are of
order 
$a^2/R_d^2$  (Appendix.~\ref{app:disc})  where
$a$ is the characteristic length of the microscopic structure (for example
a lattice constant). Therefore they are negligible in our analysis
which considers 
$R_d\gg a$.

With the above approximations the  total electric field inside the cell is 
written as 
${\bf E}={\bf E}_b+{\bf E}_d$ where $b$ ($d$) refers to the background 
(drop) contribution. 
Integrating the square of the electric field we obtain three contributions to
the electrostatic energy: $\epsilon=\epsilon_d+\epsilon_b+\epsilon_{d-b}$
with 
\begin{equation}
\label{eq:eele}
\epsilon_d=\frac1{8\pi\epsilon_0} \int d^3{\rm r} {\bf E}_d^2
\end{equation}
with $\epsilon_0$ the static dielectric constant
and a similar equation for the background. The interaction energy is
\begin{equation}
\label{eq:eelei}
\epsilon_{d-b}=\frac1{4\pi\epsilon_0} \int d^3{\bf r} {\bf E}_b.{\bf E}_d
\end{equation}
The use of the static dielectric constant is well justified because 
we are assuming a static super structure which  certainly will produce
relaxation   of the ions which in turn will screen the electric field. We are
assuming by symmetry that the electric displacement is parallel to
the electric field.
The fields can be easily evaluated with Gauss theorem. One obtains
\begin{eqnarray}
\label{eq:eeldr}
\epsilon_d&=&Q^2\frac3{5\epsilon_0 R_d} \\
\epsilon_b&=&Q^2\frac3{5\epsilon_0 R_c}\\
\epsilon_{d-b}&=&\frac{3Q^2}{\epsilon_0}
\left(-\frac1{2R_c}+\frac{R_d^2}{10R_c^3}\right)
\end{eqnarray}
where $Q\equiv - e \delta v_d$ is the effective charge inside the drop. 
The volume of a drop is $v_d=4\pi R_d^3/3$ and the number of drops is 
given by  $N_d=V_B/v_d=x V/v_d$. We also have that $x=R_d^3/R_c^3$. 
Finally the total electrostatic energy  per unit volume can be put as:
\begin{equation}
\label{eq:eelev}
e_e=\frac{2\pi e^2\delta^2}{5\epsilon_0}  R_c^2  x^{5/3} (2-3x^{1/3}+x).
\end{equation}
Setting one of the densities in $\delta$ to zero 
one recovers the expressions obtained by Nagaev 
and collaborators for the particular case of  a mixed state  
composed of  an antiferromagnetic insulating
phase and a ferromagnetic metallic phase.\cite{nag83,nag98}

The surface energy is parametrized by a quantity $\sigma$
with dimensions of energy per unit surface. In general $\sigma$
will be a function of the densities $n_A$, $n_B$. 
The total surface energy per unit volume is:
\begin{equation}
\label{eq:eeles}
e_{\sigma}=4\pi \sigma R_d^2 \frac{N_d}V=\frac{3\sigma x^{2/3} }{ R_c}
\end{equation}

These two contributions add to the mixing energy per unit volume
$e_m=e_e+e_{\sigma}$. 

Due to the constrain we have three parameters to determine ($\delta$, $x$, $R_c$). The mixing energy is the only contribution which depends explicitly on the geometry. We can therefore eliminate $R_c$ in favor of $\delta$ and $x$ by minimizing $e_m$ respect to the cell radius to get: 
\begin{equation}
\label{eq:rc}
R_c=\left(\frac{ 15\sigma\epsilon_0 }{4\pi x (2-3x^{1/3}+x)  e^{2} \delta^{2} }\right)^{1/3}
\end{equation}

Now we consider the layered geometry. The cell consist of a layer 
of width $2R_c$. The center of the cell is occupied by a layer of 
width $2R_d$ of $B$ phase and the rest is occupied by $A$ phase.
Fig.~\ref{fig:dropdr} serves as a schematic plot of the 
density profile also in this case.  $r$ is a 
coordinate perpendicular to the layers with the origin at the center
of the $B$ layer. The volume fraction now
is given by $x=R_d/R_c$. By following analogous arguments as for the drops 
we obtain:
\begin{eqnarray}
\label{eq:eelas}
e_e&=&\frac{2\pi e^2}{3\epsilon_0} \delta^2 R_c^2  x^2 (1-x)^2\\
\label{eq:eslas}
e_{\sigma}&=&\frac{\sigma}{R_c}\\
\label{eq:rclas}
R_c&=&\left(\frac{ 3 \sigma\epsilon_0 }{4\pi   x^2 (1-x)^2 e^{2} \delta^{2} }\right)^{1/3}
\end{eqnarray}

\begin{figure}[tbp]
\epsfxsize=9cm
$$
\epsfbox{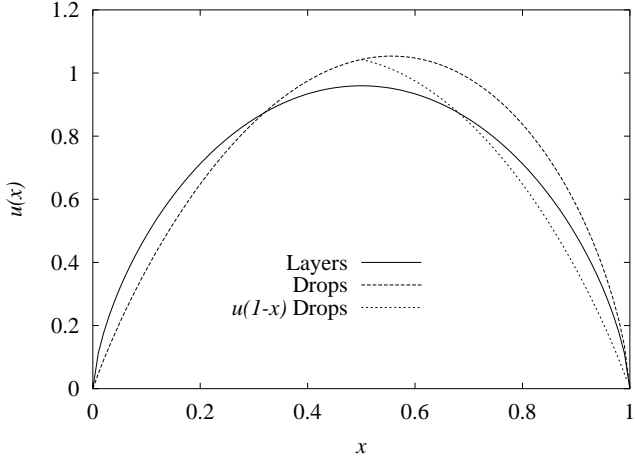} 
$$
\caption{The function $u(x)$ that parameterizes 
the mixing energy for the layer geometry and the drop geometry.
}
\label{fig:udx}
\end{figure}

Once $R_c$ has been eliminated for both geometries 
 the  mixing energy can be put as:
\begin{equation}
\label{eq:em}
e_m= \left[\frac{\sigma^2 e^2 \delta^2}{\epsilon_0}\right]^{1/3} u(x)
\end{equation}
where all the geometric information is stored in $u(x)$:
\FL
\begin{eqnarray}
\label{eq:udx}
u(x)&=&3^{5/3}\left(\frac{\pi}{10}\right)^{1/3} x
(2-3x^{1/3}+x)^{1/3} \text{\ \ \ \ (drops)} \\
u(x)&=&\left(\frac{\pi}2\right)^{1/3} [3 x (1-x)]^{2/3} 
\text{\ \ \ \ \ \ \ \ \ \ \ \ \ \ \ \ \ \ \ \ (layers)}
\end{eqnarray}
In Fig.~\ref{fig:udx} we plot $u(x)$.

The free energy 
should remain invariant  respect to an exchange of the kind
$A \leftrightarrow B$ and $x  \leftrightarrow 1-x $. 
We will term this as ``phase exchange symmetry''.  Fig.~\ref{fig:udx} 
shows that this symmetry is only approximately realized by the $u(x)$ for drops. 
 The deviation is due to the fact that the surface energy is minimized when the
 minority phase inhomogeneities are spherical. Our drop solution imposes this at 
small $x$ but violates this in the opposite case of $x\rightarrow 1$ where the 
minority phase inhomogeneities have the 
complicated geometry between the spherical drops.  In practice, however, our 
approximations 
can be meaningful even at intermediate and large $x$ because the
present $u(x)$ is approximately symmetric around $x=1/2$. This is due
to the fact that the electrostatic energy Eq.~(\ref{eq:eelev})
correctly cancels in this limit driving the total mixing energy to zero.

 A better treatment should allow at $x>1/2$ for a switch from the unoptimized 
interstitial geometry to a spherical geometry with an energy gain given by the reflected $u(x)$ at small $x$ in  $u(1-x)$ as shown in Fig.~\ref{fig:udx}. 
A comparison between the reflected curve and the original $u(x)$ shows that this 
geometry optimization compared with an apparently very bad geometry
gives rise to a modest lowering of the energy. The same happens 
when we switch from the layer geometry to the spherical drop geometry as shown in Fig.~\ref{fig:udx}.
We can conclude that the dependence on geometry is week.

The spheric drop geometry has
lower energy than the layered geometry as expected from general arguments on surface tension. The exception is close to $x=1/2$ where  our spherical
drop solution is not adequate in any case. In fact in this region drops and the crystal potential will be far from spherical. 
 The problem of establish  the optimum
geometry close to $x=1/2$ is beyond the scope of this work however
we expect minor corrections to thermodynamic quantities due to the small sensibility
of $u(x)$ to dramatic changes in the geometry as illustrated above.

Although the layer solution has higher energy due to its simplicity 
it is an excellent  test ground for checking the approximations. We take advantage
of this fact to test the UDA approximation in Sec.~\ref{sec:tf}.   
In addition the layer geometry has the extra advantage that, by construction, 
respects  the phase exchange symmetry. 

Anyway 
since $u(x)$ depends weakly on geometry  our results
for macroscopic thermodynamic quantities will be largely independent
of the geometry itself. When possible we present our
results in a geometry independent way by 
leaving  the function $u(x)$ unspecified in our expressions.

 Minimizing the free energy 
respect to $\delta$ and $x$ one obtains: 
\begin{eqnarray}
\label{eq:mu}
\mu_B-\mu_A&=&-\frac{2 (e \sigma)^{2/3} u(x)}{3(\epsilon_0\delta)^{1/3} x(1-x)}
\nonumber \\
& &+\frac{ 2 (e \delta)^{2/3} u(x) }{3 \epsilon_0^{1/3} \sigma}  
\left( \frac{1}{x}\frac{\partial\sigma}{\partial n_A} - \frac{1}{1-x}
\frac{\partial\sigma}{\partial n_B}\right) \\  
\label{eq:p} 
p_B-p_A&=& (\mu_B-\mu_A ) [n+\delta(1-2x)]\nonumber\\
&+& 
\left(\frac{e^2 \sigma^2\delta^2}{\epsilon_0}\right)^{1/3}u'(x)
\nonumber\\
&-& \frac{2\delta^{5/3} e^{2/3} u(x) }{3 (\epsilon_0\sigma)^{1/3}}
\left(\frac{\partial\sigma}{\partial n_A} 
+      \frac{\partial\sigma}{\partial n_B}\right).
\end{eqnarray}
Here $p_{A}=-f_{A}+\mu _{A}n_{A}$, ($\mu_A=\partial f_A/\partial n_A$), etc.  are the ``intrinsic''
pressures (chemical potentials) of each phase. The word
``intrinsic'' stands for the values  
of these quantities in the presence of a fictitious fully compensating
background, in other words  they  refer to a uniform single-phase 
situation. 
Equations (\ref{eq:mu}),(\ref{eq:p}) determine the jump in these 
quantities at the interface in order to have thermodynamic equilibrium
when long range Coulomb forces and surface energy are present. 
These equations are valid for a general geometry described by the function
$u(x)$. Notice that as long as $u(x)$ preserves the phase exchange
symmetry the equations also preserve this symmetry.

To analyze the effect of the long range forces and of the surface energy in 
the jumps let us neglect for simplicity the density dependence of the 
surface energy
($\partial\sigma/\partial n_A=\partial\sigma/\partial n_B=0$) and concentrate 
on the drop geometry.
Due to the different charge distributions, the electrostatic potential energy 
$-e \phi$ of an electron inside and outside the drops  is different. 
In equilibrium this jump in the electrostatic potential  should be compensated
by a jump of the intrinsic chemical potentials  [Eq.~(\ref{eq:mu})]
to make the electrochemical potential constant in  the whole system.
For $\delta >0$ the drop   repels 
electrons so the electrostatic potential energy will be lower outside the drop 
i.e. $-e \phi_A< -e \phi_B$. The intrinsic chemical potential outside will 
have to be larger than 
inside as the sign in Eq.~(\ref{eq:mu}) implies. 

Regarding the pressure, in equilibrium the intrinsic  pressure inside the
drop, $p_{B}$, should equal the pressure exerted by the host $p_{A}$
plus the pressure exerted by the mixing forces. 
For $\delta >0$   the electrostatic
energy induces a negative contribution to the pressure
since an increase in the drop volume at
constant particle number decreases the difference in densities between the
interior and the exterior of the drop and hence the Coulomb cost. 
This effect is given by the first  term  in Eq.~(\ref{eq:p}).
The second term proportional to $u'(x)$ is a geometric contribution.
Both terms are discussed in  more detail in a specific example in 
Appendix~\ref{app:ka0}.

In  the limit $e\rightarrow 0$ one gets 
$\mu _{B}=\mu _{A}=\mu $ and $p_A=p_B=p$ i.e. $\mu \delta =f_{B}-f_{A}$ 
which are the conditions for MC.

\section{General analysis of the mixed state in the uniform density 
approximation}
\label{sec:parabolic}

In this section we set up the basic ideas for  inhomogeneous
solutions. For simplicity we model each phase free energy with a parabola
and  we assume that the surface tension is
density independent. Without loss of generality we write the parabolas
as a quadratic expansion around the MC densities: 
\begin{eqnarray}  
f_A(n_A)&=&f_A^0+\mu^0 (n_A-n_A^0)+\frac{1}{2k_A}  (n_A-n_A^0)^2\nonumber\\
\label{eq:flfs}\\
f_B(n_B)&=&f_B^0+\mu^0 (n_B-n_B^0)+\frac{1}{2k_B}  (n_B-n_B^0)^2\nonumber
\end{eqnarray}
The  quantities with the ``0'' superscript (or subscript below)
satisfy MC in the absence of LRC forces i.e. 
$f_B^0-f_A^0=\mu^0 \delta_0$ and $\delta_0=n_B^0-n_A^0$. 
The linear sloop $\mu^0$ is the same for the two phases due to the MC 
condition. 
The MC  density $n_0$ and the volume fraction are related
by  $n_0=n_A^0+\delta_0 x$. The constants $k_A$, $k_B$ are essentially the
compressibilities of the two phases.\cite{note}

For non interacting electrons at T=0 the  compressibility 
coincide with the density of states. For the 3D free electron gas we have: 
\begin{equation}
\label{eq:kfree}
k_{free}=\frac{3^{1/3}m n_0^{1/3}}{\pi^{4/3} \hbar^2}
\end{equation}
with $m$ the electronic mass. 


 Another useful realization is  a nondegenerate gas where we have:
\begin{equation}
\label{eq:gas}
k_{gas}=\frac{n_0}{KT} \label{eq:kgas}
\end{equation}

Our aim in the following is to 
obtain the equations which control  the deviation from MC
behavior in the presence of the mixing energy.

We define a dimensionless global density 
$$n'\equiv(n-n_A^0)/\delta_0$$
which measures the distance from the point in which $B$ phase appears 
in the absence of Coulomb forces. In MC the coexistence
region is given by  $0<n'<1$.  

Eqs.~(\ref{eq:mu}),(\ref{eq:p}) determine $\delta$, and $x$ for a 
fixed density where now $\mu_A$, $\mu_B$, $p_A$, and $p_B$ can be expressed
in terms of the parameters appearing in Eqs.~(\ref{eq:flfs}).

In practice it is much easier to solve the equations by
fixing the volume fraction $x$ and solving for $\delta$,  and $n$,
i.e. we find which density one should put in the system to obtain a mixed state
with a  given volume fraction. This is because the solutions happen to be 
multivalued functions of $n$ whereas they are single valued  functions of $x$
(see below). 

For a fixed volume fraction $x$ we define the dimensionless density 
deviations from the MC values: $\hat n= (n-n_0)/\delta_0$ and 
$\hat \delta=  (\delta-\delta_0)/\delta_0$.
The density deviation  $\hat n$ measures the shift in the global 
density needed to have the same volume fraction of a system without 
LRC interaction. 

To fix the energy units it is convenient to choose one of the two 
compressibilities as a reference, for example the largest. We define
$k_m=\max(k_A,k_B)$. Energies per unit volume will be measured in units of
the  characteristic PS energy $\delta_0^2/k_m$. 
The latter  is essentially the difference  
between the uniform parabolic free energy and the MC free energy at the 
characteristic density $\delta_0$.

Now we define two important reference lengths scales in the theory.  
The characteristic size of an inhomogeneity
 for which the Coulomb energy balance the 
surface energy is given by the $R_c$ of previous section
 with the geometric factors dropped
and the density evaluated at the MC value.  This define the scale  
\begin{equation}
  \label{eq:ld}
l_d=\left(\frac{  \sigma\epsilon_0 }{ e^{2} \delta_0^2 }\right)^{1/3}.  
\end{equation}
 The other length  is given by $l_s^2=\epsilon_0/(4 \pi e^2 k_m)$.
By relaxing the UDA we will show in section Sec.~\ref{sec:lda} that 
$l_s$  is a screening length. In other words if the reference phase 
(the one with $k_m$) is interpreted as a metal $l_s$ is the
characteristic distance in which the  electric field penetrates.

The theory has two dimensionless  parameters. One is the ratio $k_B/k_A$. 
The other  measures the strength of
the mixing energy energy effects in units of the characteristic PS energy
$\delta_0^2/k_m$ and is given by:
\begin{equation}
\label{eq:la}
\lambda=2 \frac{k_m}{\delta_0^2}
\left(\frac{9\pi e^2\delta_0^2\sigma^2}{5\epsilon_0  }\right)^{1/3}= 
\frac12\left(\frac{9}{5\pi^2} \right)^{1/3}
\left(\frac{l_d}{l_s}\right)^2  
\end{equation}
The characteristic mixing  energy is given by the factor  with the 
power $1/3$ in the middle expression. The constant 
$\lambda$ characterizes the competition of the mixing energy 
cost and the MC like energy gain due to phase separation. The
coupling  constant goes to zero
as  $e\rightarrow 0$ with $\sigma$ finite. This correspond to the usual 
PS. The case   $\sigma\rightarrow 0 $ with finite $e$
correspond to an infinite number of drops (or layers) of zero radius. 
In this maximum intermixing
situation the charges of the two phases spatially  coincide and the Coulomb 
cost goes also to zero so that the MC is again valid. Notice however 
that this last idealized situation
cannot be reached in practice because at some point for small drop radius 
the continuous approximation will fail. 

Inserting the explicit expressions [Eqs.~(\ref{eq:flfs})] of $f_A$ and $f_B$
in Eqs.~(\ref{eq:mu}),(\ref{eq:p}) we obtain the 
following equations for the density deviations:
\FL
\begin{eqnarray}
\label{eq:emuep}
\hat n \left(\frac{1}{k_B}-\frac{1}{k_A}\right) +
\hat \delta  \left(\frac{1-x}{k_B}+\frac{x}{k_A}\right)=\nonumber \\
\left(\frac{5}{9\pi}\right)^{1/3} \frac{\lambda u(x)} {3 k_m 
(1+\hat\delta)^{1/3}x(1-x)}\nonumber \\
\\
\frac{ x \hat \delta -\hat n }{k_A}   +
\left[\hat n \hat \delta (1-x)+
\frac{{\hat n}^2}2 \right] \left(\frac{1}{k_B}-\frac{1}{k_A}\right)+\nonumber\\
\frac{{\hat \delta}^2}2  
\left[\frac{1-2 x}{k_B}+\frac{2 x}{k_A}+ 
\left(\frac{1}{k_B}-\frac{1}{k_A}\right) x^2\right]=\nonumber\\
\left(\frac{5}{9\pi}\right)^{1/3} \frac{ \lambda (1+\hat
\delta)^{2/3} }{2k_m} \left[ u'(x)+\frac{2u(x)}{3 (1-x)} \right] . \nonumber
\end{eqnarray}

Eqs.~(\ref{eq:emuep}) can be solved numerically for general values of the
parameters. For small $\lambda$ i.e.  for small mixing energy  we can 
linearize the equations neglecting higher order terms in  $\hat \delta$ and 
$\hat n$.  We will refer to this as the linearized UDA. We get:
\begin{eqnarray}
\label{eq:emuepl}
\hat n \left(\frac{1}{k_B}-\frac{1}{k_A}\right) +
\hat \delta
\left(\frac{1-x}{k_B}+\frac{x}{k_A}\right)=\nonumber\\
\left(\frac{5}{9\pi}\right)^{1/3} \frac{\lambda u(x)} {3 k_m x(1-x)}
\\
\frac{ x \hat \delta -\hat n }{k_A} 
=- \left(\frac{5}{9\pi}\right)^{1/3} 
\frac{\lambda}{2k_m}\left[ u'(x)+\frac{2u(x)}{3 (1-x)} \right]
\end{eqnarray}
For the sake of simplicity in the following we will consider the 
linearized solution. We checked that  for all the physical properties the 
difference between the linearized and the exact solution is quite small 
in the range of $\lambda$ where the drop solution is stable.

The linearized solution takes a simple form and is explicitly
symmetric respect to an exchange of phases when written in the original 
variables:
\begin{eqnarray}
n_A&=&n_A^0+\frac16 \left(\frac{15}{\pi}\right)^{1/3}   \frac{k_A}{k_m} \lambda \delta_0 \left[ u'(x)+\frac{2u(x)}{3 (1-x)} \right] \nonumber   \\
& &\label{eq:solnlns}\\
n_B&=&n_B^0+\frac16  \left(\frac{15}{\pi}\right)^{1/3}   \frac{k_B}{k_m}
\lambda \delta_0  \left[ u'(x)-\frac{2u(x)}{3 x} \right].  \nonumber
\end{eqnarray}
In the case of $\lambda=0$, according to MC, the system separates
in two phases  with densities $n_A^0$,  $n_B^0$ respectively
independently of the volume fraction.
For nonzero $\lambda$ 
and small $x$ the $B$ phase divides in drops  or layers and the density in each
phase depends on the volume fraction of $B$ phase. The deviation of
each density 
from MC prediction is proportional to $\lambda$ and to the
compressibility of each phase.  Notice that the density of an 
incompressible phase ($k\rightarrow 0$) does not depend on the volume
fraction even in the presence of LRC forces.

\begin{figure}[tbp]
\epsfverbosetrue
\epsfxsize=9cm
$$
\epsfbox{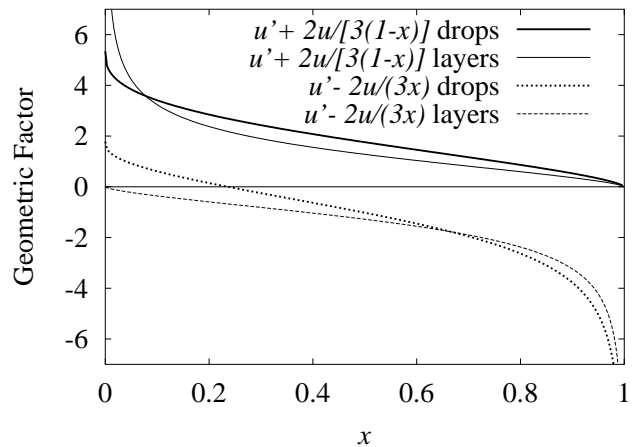}
$$
\caption{The dimensionless functions that determine the change in
$n_A$ (upper curves) and $n_B$ (lower curves) for small $\lambda$
vs. the volume fraction $x$ for the layered and the drop geometry.
[See Eq.~(\protect\ref{eq:solnlns})]  }
\label{fig:nsnldx}
\end{figure}

In Fig.~\ref{fig:nsnldx} we show the behavior of the two functions 
which determine the dependence of the densities on the volume fraction. 
In the drop geometry and for small $x$ both  $n_A$ and  $n_B$ 
tend to be larger than in the MC case whereas in the layered geometry
only $n_A$ is larger. This gives rise to minor qualitative differences
in the behavior of drops and layers. Apart from this  the overall
behavior is similar.  

The equation for the density of one phase [Eq.~(\ref{eq:solnlns})] has  
a transparent interpretation in the limit in which the 
other phase, say $A$, is incompressible ($k_A=0$). 
This case is solved in detail in Appendix~\ref{app:ka0}.
Assume that $A$ phase is the vacuum and so exerts no pressure and has 
zero density. We can consider that the mixing forces  due to the 
 electrostatic and surface energies exert an  
 ``external'' pressure  on the $B$ phase inhomogeneity. 
  In equilibrium the intrinsic pressure of $B$ phase 
($p_B$)  should compensate this ``mixing pressure'' ($p_B=p_m$). 
The latter is  shown in Appendix~\ref{app:ka0} to be  given by:
\begin{equation}
\label{eq:pm}
p_m= \frac{\partial e_m}{\partial x}-  \frac{2e_m}{3 x}
\end{equation}
On the other hand a change in the external pressure correspond 
to a change in the $n_B$ density according to  the 
$B$ phase equation of state.    This follows directly from our definition
of  compressibility:\cite{note}
$$
k_B\equiv n_B^0 \frac{\Delta n_B} {\Delta p_B}
$$
where we have replaced a derivative by a finite different ratio. 
We can obtain the second linearized expression in 
Eq.~(\ref{eq:solnlns}) 
directly  from this definition using that 
the MC density correspond to zero intrinsic pressure.
$$n_B-n_B^0=k_B  \frac{p_B}{n_B^0} 
\propto k_B n_B^0 \left[ u'(x)-\frac{2u(x)}{3x} \right].$$ 

The mixing pressure can be negative as explained in the 
Appendix~\ref{app:ka0}. This implies that the 
density is less that the MC density. From the lower cures in 
Fig.~\ref{fig:nsnldx} we see that 
for drops the pressure is positive for small $x$ and then becomes negative
whereas for layers the pressure is negative for all $x$. 

Remarkably in both cases the mixing pressure is a decreasing function of $x$.
Since in general  $x$ is an increasing function of $n'$  we can anticipate 
that $n_B$ will decrease as $n'$ increase (see below).
Notice that for small $x$ we have $p_B\sim u'(x)/3$ so a decreasing 
mixing pressure can be directly related to the negative curvature of $u(x)$ 
(Fig.~\ref{fig:udx}).

Coming back to the general solution in Eq.~(\ref{eq:solnlns}) 
we are interested in the dependence of these quantities  
as a function of the global density $n'$, our true control variable, 
rather than as a 
function of the volume fraction. Hence we need  the volume fraction as a 
function of the global density $n'$. From the solution of the linearized 
equations we find:
\FL
\begin{eqnarray}
\label{eq:ndx}
&&n'=x +\left(\frac{15}{\pi}\right)^{1/3} \frac{\lambda}6 \times\\
&& \left(\frac{k_A (1-x) }{k_m}   \left[ u'(x)+\frac{2u(x)}{3 (1-x)} \right]
+  \frac{k_B x} {k_m  }     \left[ u'(x)-\frac{2u(x)}{3 x} \right] \right) \nonumber 
\end{eqnarray}
Since all physical quantities depend on the densities 
this completes the solution of the problem.  

Specific results will be presented in the next section for the drop
geometry and in Sec.~\ref{sec:tf} for the layered geometry. 


\subsection{Results of the UDA for the drop geometry}
\label{sec:uda}

Now we consider the drop geometry and 
we analyze in detail the two cases: i)  the
compressibilities of the two phases are equal ($k_B=k_A=k_m$)
and ii) one of the compressibilities is zero.

\begin{figure}[tbp]
{\hspace{0.cm}{\psfig{figure=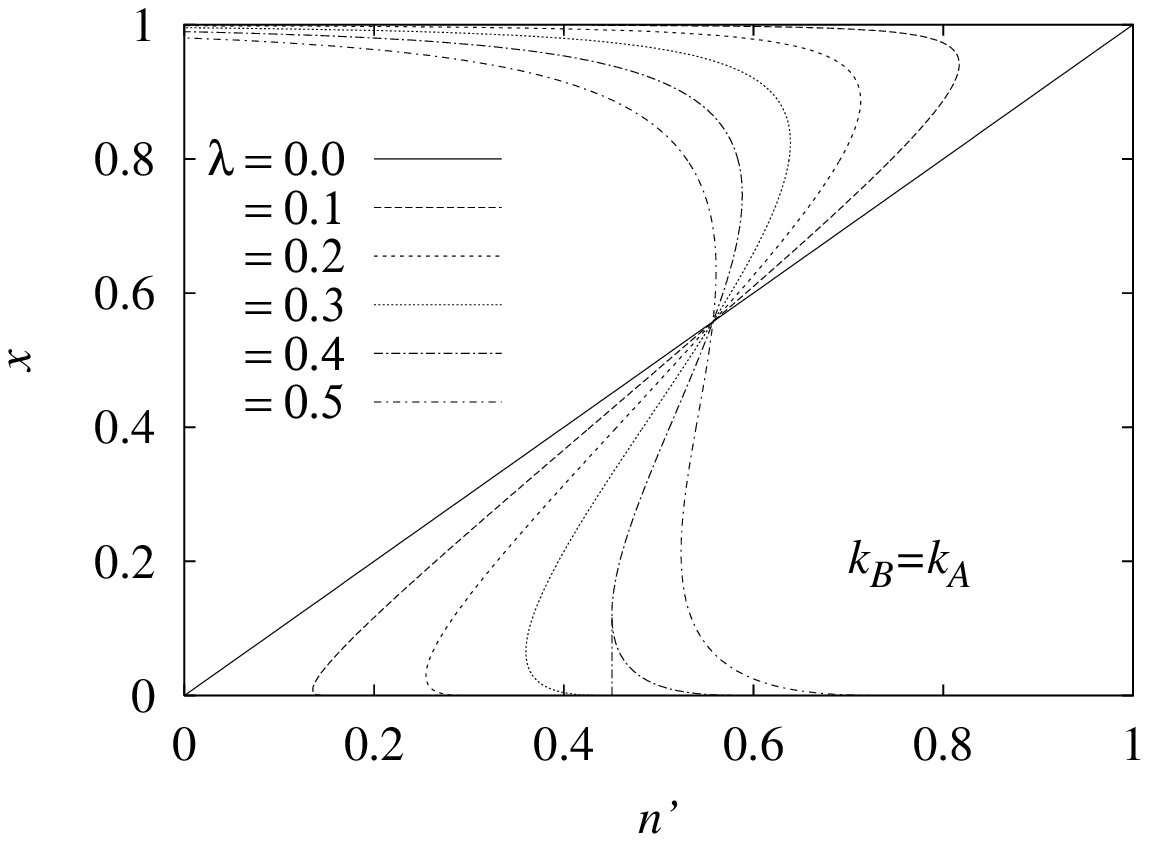,width=9cm}}}
{\hspace{0.cm}{\psfig{figure=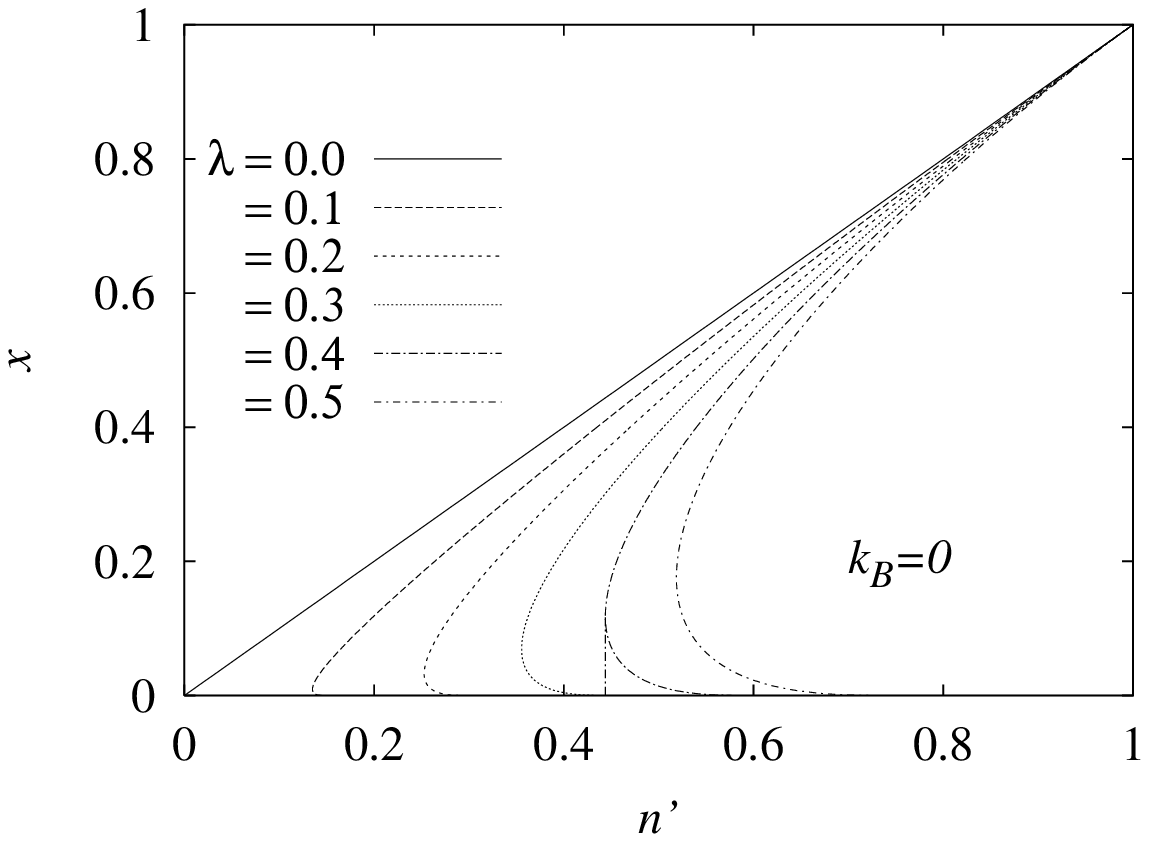,width=9cm}}}
\caption{Top panel: Volume fraction vs. $n'$ for 
(from left to right at the bottom) 
$\lambda=0,0.1,0.2,0.3,0.4,0.5$ and  $k_B=k_A$. For $\lambda=0.4$ we 
indicate with a
vertical line the discontinuity in the volume fraction to go 
from the uniform solution to the drop solution by increasing the density.
Bottom panel: Same for  $k_B=0$.
The approximations done are rigorously valid only for small $x$.
}
\label{fig:xdn}
\end{figure}

In Fig.~\ref{fig:xdn} we plot the volume fraction as a function of 
global density from Eq.~(\ref{eq:ndx}) for the drop solution.
The volume fraction is a multivalued function of $n'$ and 
in the case $k_B=k_A$ has 
a lower branch close to $x=0$, an intermediate branch, and  an upper 
branch close to $x=1$. 
The intermediate branch is the physical solution. This will be shown below
by looking at the free energy. 
The physical solution   has the intuitive  property that the volume 
fraction increases as global density increases.

We see that the  bifurcation density  $n_{bif}'$  at which the phase 
separated solution appears for finite $\lambda$ is larger than in MC.
On the other hand  $B$ phase  appears with   a finite volume 
fraction and  its growing  rate is larger than in the MC case.
Remarkably  both the volume fraction at the bifurcation point and the
bifurcation density $n_{bif}'$ are almost the same for 
$k_B=k_A$ and for $k_B=0$. They depend only on $\lambda$ as can be seen 
by comparing  the two panels in Fig.~\ref{fig:xdn}.

In the case $k_B=0$ the constraint between the volume fraction and the 
densities  together with the fact that the $B$ density is fixed make all 
the curves 
to converge to the MC case when $x\rightarrow 1$ as shown in
Fig.~\ref{fig:xdn}. The same happens 
when $k_A=0$ and  $x\rightarrow 0$.

\begin{figure}[tbp]
\epsfxsize=9cm
$$
\epsfbox{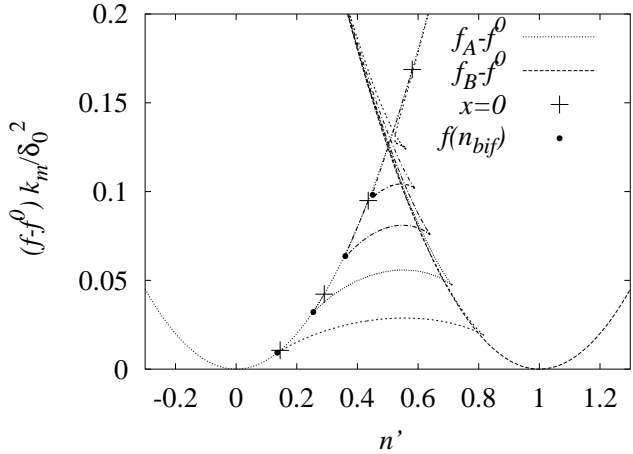} 
$$
\caption{
$f_A-f^0$, $f_B-f^0$, and $f-f^0$ in the drop solution  
for   $\lambda=0.1,0.2,0.3,0.4,0.5$ (from bottom to top) and $k_B=k_A$ 
vs.  $n'$. Here 
$f^0$ is the MC free energy for $\lambda=0$  (a straight line). The
cross indicates the value with $x=0$ of the drop solution for each $\lambda$.  
The black dot indicates the bifurcation point in which the drop 
solution first appears when density increases. 
}
\label{fig:fdn}
\end{figure}

\begin{figure}[tbp]
\epsfxsize=9cm
$$
\epsfbox{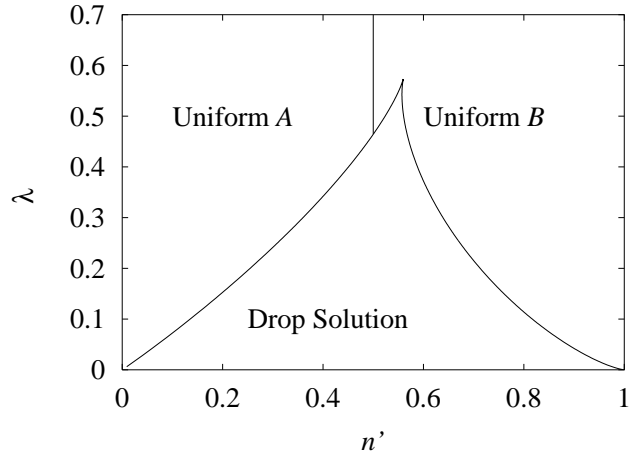}
$$
\caption{Locus of existence of the low-energy drop solution in the $\lambda$-$n'$ plane
for $k_A=k_B$.
This almost 
coincides with the phase digram in the sense that when the drop solution 
exist it is more stable than the uniform solution except close to 
$\lambda=0.5$ and
in a very narrow region around the drop-uniform boundary line (see text). }
\label{fig:l_bifdn}
\end{figure}

To decide the stability of the solution we have to compare the 
drop solution with the single phase solution. 
In Fig.~\ref{fig:fdn} we show $f_A(n')$, $f_B(n')$ and  the total
free energy with $k_B=k_A$
for various $\lambda$. The   MC line $f^0(n')=f_A^0+n'(f_B^0-f_A^0)$
has been subtracted. 
The energy also is a multivalued function of $n'$.
As the density increases the drop solution appears  at $n_{bif}'$ (indicated 
in Fig.~\ref{fig:fdn} by a black dot) 
with two different branches.
In the upper (unstable) branch  $x$ decreases with 
density till the point $x=0$ highlighted with a cross in Fig.~\ref{fig:fdn}.
For the lower branch one finds the expected behavior i.e. $x$ increases
with density. The upper branch is almost degenerate with the bulk 
$f_A(n)$ free energy. Near the bifurcation the three solutions (homogeneous,
drop stable and drop unstable) are very close in energy. Approximation in
the solution of the Eq.~(\ref{eq:emuep}) can  lead to wrong 
conclusions about the relative stability. In this case one has to refer 
to the non-linearized solution. For the latter (not shown) 
we find that the bifurcation
density $n_{bif}$ is lower than the density $n_c$ at which the energy of the
lower energy drop solution crosses the energy of the uniform phase $f_A(n')$. 
However the difference between $n_c$ 
and  $n_{bif}$ is negligible for all practical proposes except for the largest
$\lambda$. In this case there is a small region 
($\lambda_c=0.49< \lambda< 0.57$) 
in which the lower energy drop solution still exist but is less stable than 
the  homogeneous solution. 
If we neglect this small effect the phase diagram 
of the drop solution is given by  $n_{bif}$ vs. $\lambda$ as shown 
in Fig.~\ref{fig:l_bifdn}. 
The uniform-drop boundary line is determined by 
the condition 
$\partial n'/\partial x=0$ (see Fig.~\ref{fig:xdn}).

For $\lambda> \lambda_c$   the homogeneous solution is stable for 
any global density. The uniform A-B boundary line is determined 
by the crossings of the parabolas in Fig.~\ref{fig:fdn}. 

When one of the compressibilities goes to zero, say $k_B$, the crossing moves 
to the right in  Fig.~\ref{fig:fdn} and the uniform $B$ region shrinks until
the boundary line for uniform B phase approaches the MC value ($n=n_B^0$).
At the same time $\lambda_c$ increases.
Analogous results are obtain for $k_A$ going to zero.  

\begin{figure}[tbp]
\epsfxsize=9cm
{\hspace{0.cm}{\psfig{figure=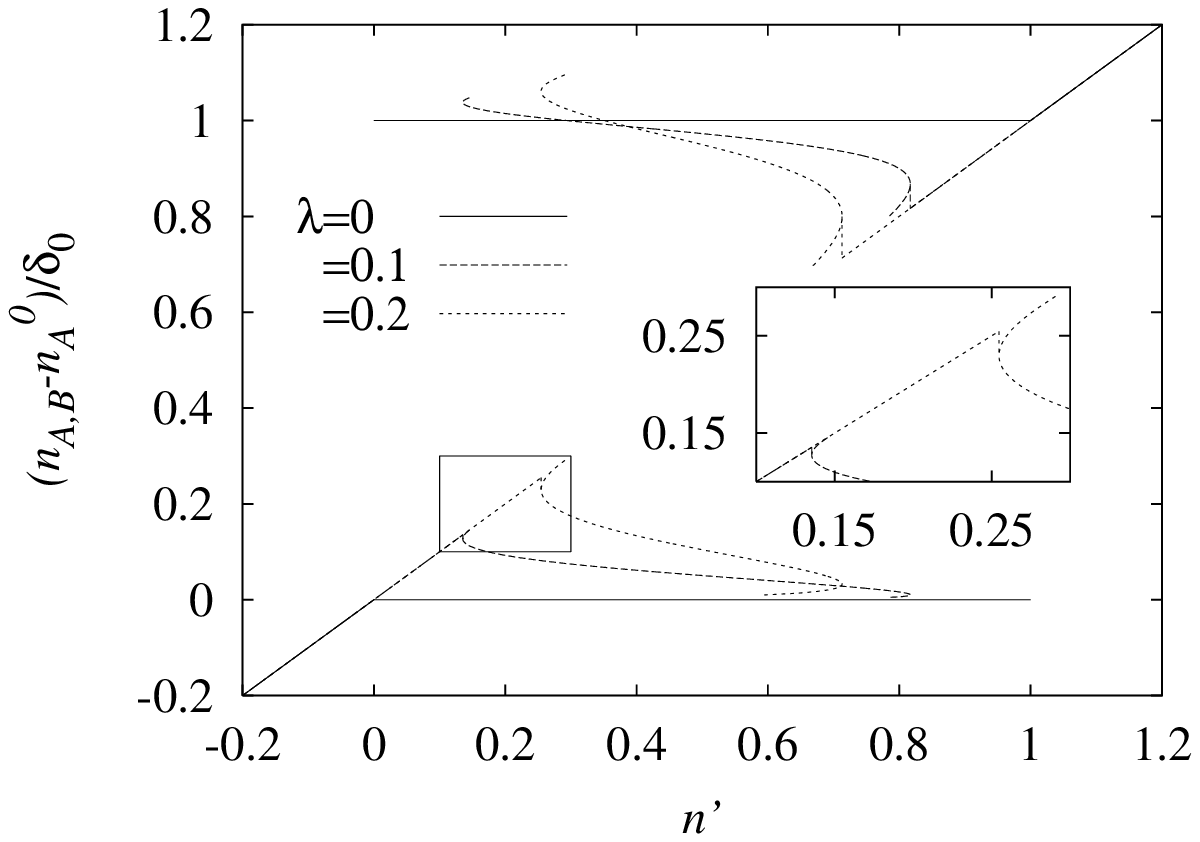,width=9cm}}}
{\hspace{0.cm}{\psfig{figure=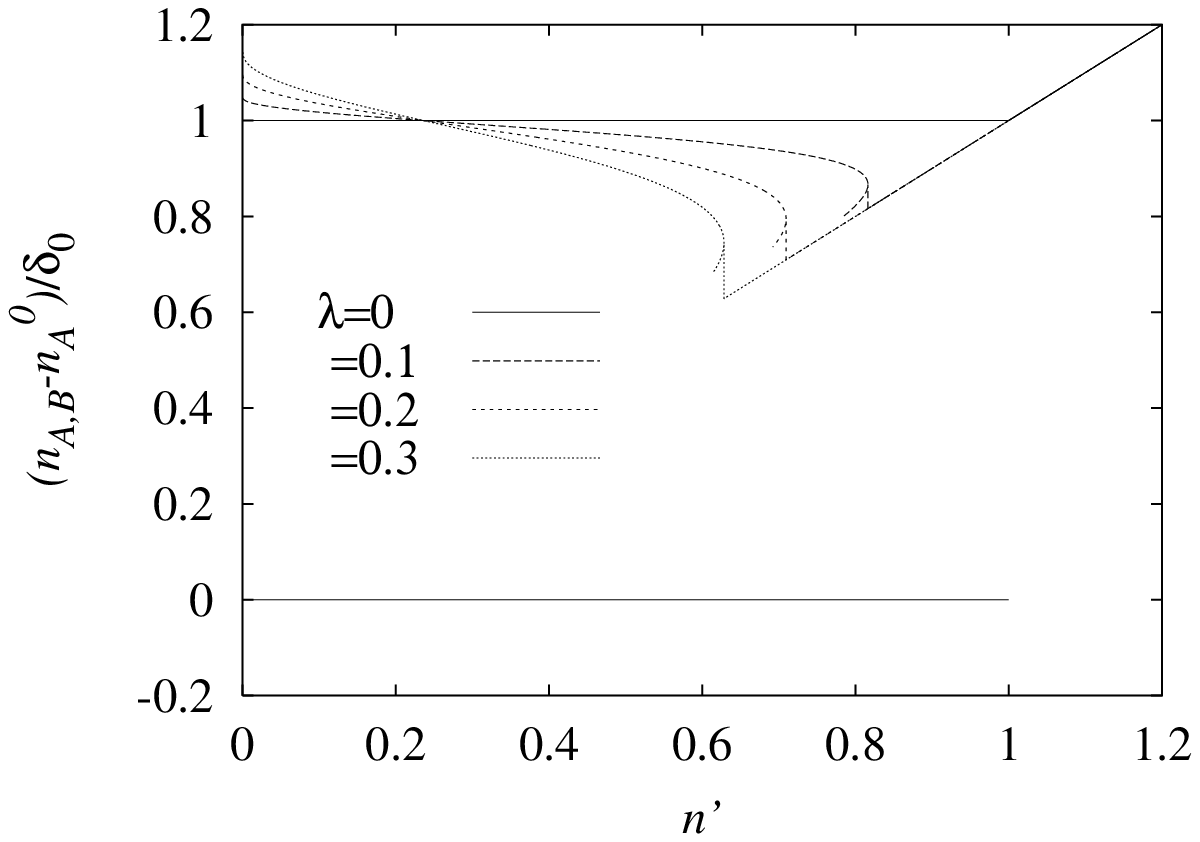,width=9cm}}}
\caption{Normalized densities of each phase as a function of normalized 
global density $n'$ for different $\lambda$. 
The upper panel is for $k_B=k_A$ and the lower panel is for
$k_A=0$. 
For each panel the lower curves  correspond to $A$ phase and the 
upper curves 
to $B$ phase. In the coexistence region multivalued densities 
appear. The long branch is the physical one and  
the short branches are unphysical.
The inset shows an enlargement of the $A$ density to resolve the 
discontinuities.   }
\label{fig:nsnldn}
\end{figure}

In the upper panel of Fig.~\ref{fig:nsnldn} we show the density of each 
phase as a function of the global densities for $k_B=k_A$. Increasing the 
global density the transition occurs from the uniform $A$ phase,
with density higher that the MC one,  to the 
drop state. In the MC case the density of $A$ phase is continuous 
at the transition and remains constant in the coexistence region.
For nonzero $\lambda$ the $A$ density has a discontinuity 
when the drops occur (see inset).  Remarkably both local densities decrease
as the global density increases. This is due to the behavior of the 
mixing pressure as explained above and 
in Appendix~\ref{app:ka0}. In the  case $k_A=0$ the regions whit
$n_B>n_B^0$ ($n_B<n_B^0$) can be directly associated with positive 
(negative) mixing pressures. 

Compared to the upper panel 
the lower curves for $n_A$ shrink to the MC case and the upper curves for
$n_B$ remain very 
similar (even quantitatively) except close to $n' \rightarrow 0$.
We mention that in the case  $k_B=0$ (not shown) a similar effect is 
seen exchanging  $A$ with  $B$.

\begin{figure}[tbp]
\epsfxsize=9cm
{\hspace{0.cm}{\psfig{figure=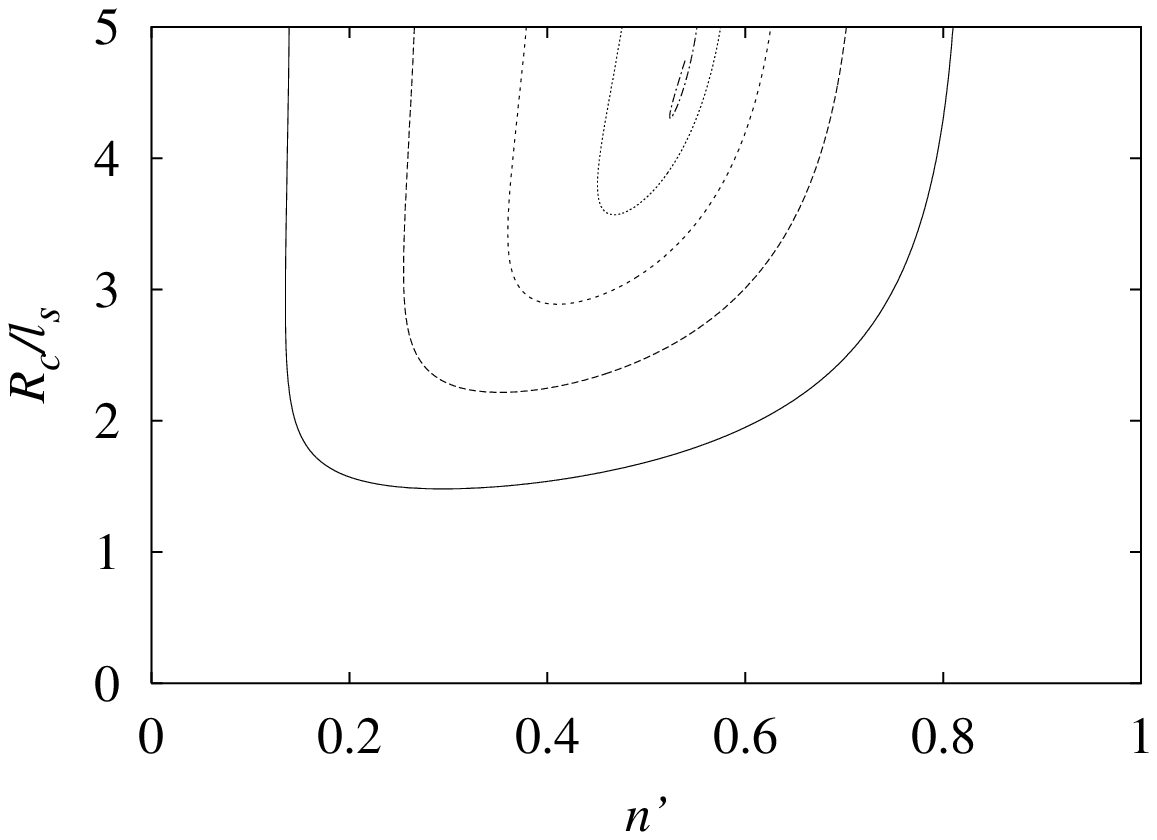,width=9cm}}}
{\hspace{0.cm}{\psfig{figure=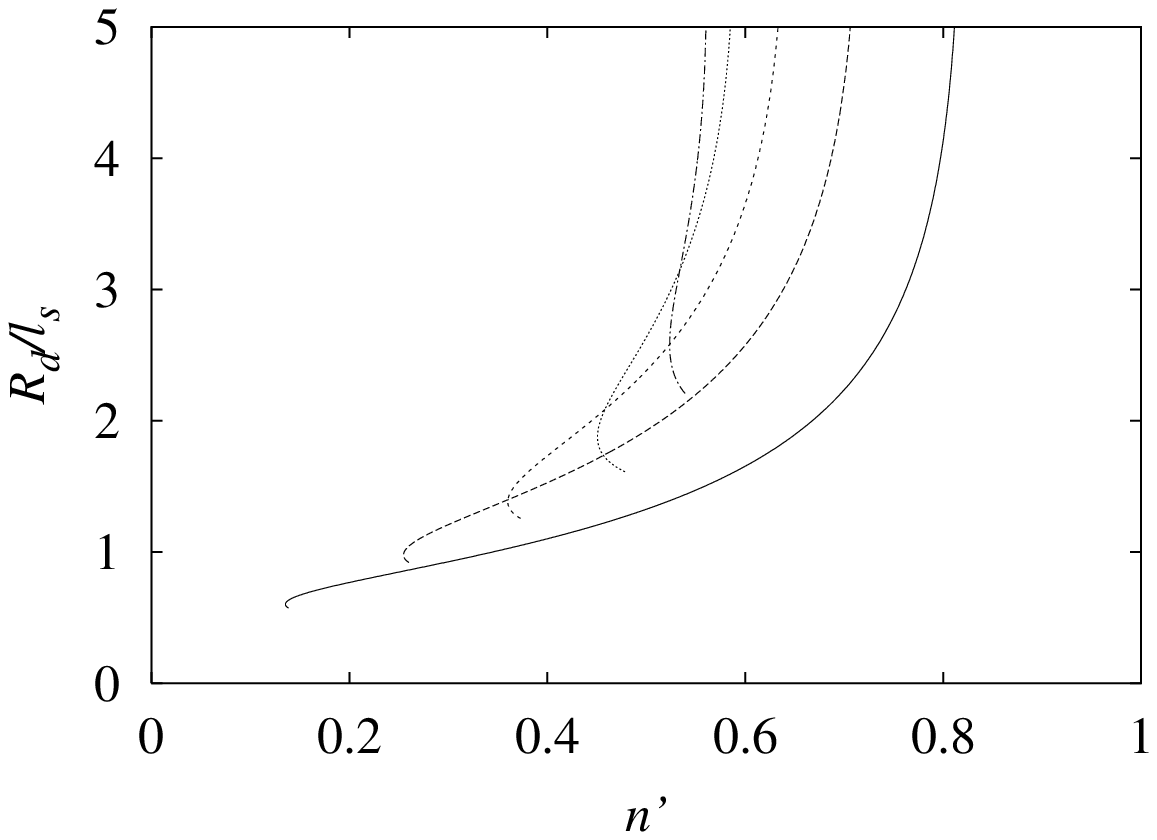,width=9cm}}}
\caption{
$R_c$ and  $R_d$ in units of the screening length  $l_s$
defined above  Eq.~(\protect\ref{eq:la}) vs.  $n'$ for $k_B=k_A$. We show the 
curves for   $\lambda=0.1,0.2,0.3,0.4,0.5$
which  increases from  bottom to top in the top panel and from 
right to left at the top in the lower panel.
In the top (bottom) panel for each curve the lower (upper) branch is
the stable one.}  
\label{fig:rcdn}
\end{figure}

In Fig.~\ref{fig:rcdn} we show the cell radius and drop radius 
in units of the screening length as a function of density for $k_B=k_A$. 
Both the cell and the drop radius  are  typically on the scale of a
few screening lengths $l_s$ for not too small $\lambda$ 
and have a finite
size at the appearance of the mixed state. The cell radius  decreases as the 
density goes away from the bifurcation value to reach a
minimum  close to $n'=1/2$. 
The minimum would be exactly at $n'=1/2$
in an exact computation  due to phase exchange symmetry. This is 
show below in the layered solution. 

The ($B$ phase) drop radius 
instead is intrinsically asymmetric and increase monotonously
with the density reflecting  the transformation of the cell from
$A$ phase to $B$ phase.

For 
$\lambda \rightarrow 0$ the cell radius and the drop radius  behave  as 
$R \sim \sqrt{\lambda} l_s \sim [\sigma\epsilon_0/(\delta_0 e)^2]^{1/3}$.
As stated in Sec.~\ref{sec:freee} 
they  diverge as $e \rightarrow 0$ 
indicating that MC can be realized with a single large drop of $B$
phase in $A$.

Another peculiarity of the curves in Fig.~\ref{fig:fdn} is that the
free energy of the drop solution has the ``wrong'' curvature, that is
the compressibility (defined from $\partial^2f/\partial^2 n$) is negative. 
This does not necessarily imply an 
instability since the usual  stability condition of positive
compressibility is formulated for a neutral system, that is including
the background compressibility.  Since we are assuming the inverse 
background compressibility to be an infinite positive number
(in our analysis the background density has a fixed homogeneous value) 
it follows that 
the total compressibility is positive and from this point of view the system
is in a stable mixed state.   Of course this does not guarantee
stability against more complicated solutions than the simple crystal of drops. 

The situation is more severe for $\lambda>\lambda_c$ where the drop solution, 
if it exists, is not stable.  In this case,
the system remains always single phase and the free energy is given by the
branches of the 
parabola with the smaller energy in Fig.~\ref{fig:fdn}. It    
changes suddenly from $A$ phase to $B$ phase at the density
$n'_c$ for which $f_A(n'_c)=f_B(n'_c)$. (For our parameters $n'_c=0.5$).
The problem is now that the energy has a cusp pointing
upwards at $n'_c$ which implies an infinite negative inverse compressibility.
This will compete with the infinite positive inverse compressibility of the
background.   Clearly one should consider in this case the background 
compressibility (e.g. the lattice compressibility) since the beginning. 
As a first step we can add to the above electronic free energy a 
background  free energy contribution $f_b(n)=(n-n_c)^2/{2k_b}$. A very rigid
(but not infinitely rigid) background is described by a very small 
$k_b>0$ which correspond to a  very narrow parabola for the background free energy. 
The total free energy, background plus cusp, will have a cusp  pointing up with two local
minima nearby. Since now the total free energy corresponds to a truly neutral system one 
can make a MC between the two local minima. One obtains a phase 
separation between $A$ and $B$ with the background adjusting its density 
in each region to the density of each phase to make it neutral.
The same argument applies at the critical density where the drop solution crosses the
uniform solution, although the negative dip is much less pronounced in that case.  Usually the electronic system
is a crystal where the background is provided by the ionic lattice.
If one trays to prepare the crystal with an electronic density
close to the critical one the system can break in two pieces each one with 
a different lattice constant. Typically the crystal is not at a fixed volume
but at a fixed external pressure $P$. (We use capital $P$ to distinguish the 
pressure exerted on the crystal as a whole from the electronic pressures 
of the phases $p_A$ and $p_B$). 
In this situation MC determines the equilibrium pressure $P_0$ for phase 
coexistence. $P_0$ will depend on the global doping so above 
$\lambda_c$, $P_0$ vs. doping determines a phase boundary line which will cut
ambient pressure at some critical doping.   

Since the electronic free energies  depend on external parameters,
a remarkable implication  is that   the critical doping will also
depend on external parameters like magnetic field, temperature, pressure, 
etc. In other words a crystal can be driven from a single phase to a 
two phase situation by changing external parameters. This is very reminiscent
of the situation in some manganites where one finds that a single phase crystal
brakes in a multidomain crystal by lowering the temperature. The multidomain
system shows lattice 
mismatch and large stress at the interfaces.\cite{ueh99,cox98} 

In Sec.~\ref{sec:tf} analogous  results  are
presented for the layered geometry case and compared with a more elaborate
computation which relax the UDA. In paper 
II we apply to different physical systems the ideas developed in 
this section. 

\section{Local density approximation}
\label{sec:lda}
In this section we generalize our results to take into account the 
full spatial dependence of the density. The basic assumption is that
we can write the free energy of each phase as the spatial integral of a  free
energy density which is a function of the local density. i.e. we are
using a local density approximation (LDA). The free energy reads:
\begin{eqnarray}
\label{eq:fdndr}
F&=& \int_{{\bf r}\in A} d^3{\bf r}  f_A[n({\bf r})]+
\int_{{\bf r}\in B} d^3{\bf r}  f_B[n({\bf r})]  \nonumber\\ 
&+& \frac{1}{  8 \pi}  \int  d^3{\bf r} {\bf E}^2 +   S_{AB} \sigma  
\end{eqnarray}
Here ${\bf r}\in A$ indicates that the integral is restricted to the
regions of phase $A$ and  $S_{AB}$ is the total interface surface
between $A$ and $B$ and we assume for simplicity $\epsilon_0=1$. 
One should be careful  not to double count in
$\sigma$ surface energy costs that are due to the spatial variation of 
the charge since this will be explicitly taken into account in the 
first three  terms. On the other hand one can include in  $\sigma$ other
effects, like magnetic, which would not be included otherwise. 
For simplicity we will assume $\sigma$ to be density independent. 

The electric field is related to the total charge density (electronic plus
background) through 
Poisson equation:
\begin{equation}
\label{eq:poisson}
{\bf \nabla . E}= 4 \pi \rho
\end{equation}
with the total charge density:
\begin{equation}
\label{eq:rho}
\rho= -e [n({\bf r})-\bar n]
\end{equation}
Here $\bar n$ is the global density of the previous section and the  
bar  distinguishes it from the spatially varying
density $n({\bf r})$. Notice that $e \bar n$ is the charge density of the 
background. The condition of neutrality is written as:
\begin{equation}
\label{eq:neut}
\bar n=\frac1V \int_{{\bf r}\in A} d^3{\bf r}  n({\bf r} )+ \frac1V
\int_{{\bf r}\in B} d^3{\bf r} n({\bf r})  
\end{equation}
Using   $n({\bf r})=n_A$ for ${\bf r}\in A$ and $n({\bf r})=n_B$ for 
${\bf r}\in B$ one recovers the UDA. 

Instead of minimizing the functional with respect to the density it is
convenient to use Eqs.~(\ref{eq:poisson}),(\ref{eq:rho}) to express the
density as a function of the electric field,  $[n= n({\bf \nabla . E})]$
and minimize the functional with
respect to the electric field profile. We look for periodic solutions
(layer, crystal, etc) and restrict the computation 
to one cell.

Minimizing the free energy [Eq.~(\ref{eq:fdndr})] respect to the electric 
field one obtains:
\begin{equation}
\label{eq:edf}
{\bf E}= -\frac1e \nabla\frac{\partial f_{X}}{\partial n}[n({\bf \nabla . E})]
\end{equation}
where $X=A$ or $B$ when ${\bf r}\in A$ or ${\bf r}\in B$ respectively.
This differential equation together with the boundary condition
determines the field profile. The boundary condition at the cell boundary 
and at the internal boundary will be discuss in the example below. 
Once the electric field profile is known for a given geometry
 the density profile is given by Poisson equation. As a final step one
  should optimize the geometry. 

Introducing the parabolic expressions [Eqs.~(\ref{eq:flfs})]
 to parameterize the  free
energy densities in Eq.~(\ref{eq:edf}) one obtains:
\begin{equation}
\label{eq:ede}
{\bf E}= l_X^2 {\bf \nabla}{\bf \nabla . E}
\end{equation}
with $l_X^2=(4 \pi e^2 k_X)^{-1}$. Clearly $l_X$ is the screening length as
anticipated in Sec.~\ref{sec:parabolic}. If we use the compressibility
of a free electron gas for $k_X$  [Eq.~\ref{eq:kfree}]
and reintroduce the dielectric constant $l_X$  corresponds to the
Thomas-Fermi screening length:
\begin{equation}
\label{eq:ltf}
l_{X}^2=\left(\frac{\pi}3\right)^{1/3} 
\frac{\epsilon_0 \hbar^2}{4 e^2  m (n_X^0)^{1/3}}.
\end{equation}
We reach Thomas-Fermi theory
which is the simplest version  of the LDA used for  electronic
structure computations. If we use the nondegenerate gas
compressibility [Eq.~(\ref{eq:kgas})]
$l_X$ is the
Debye-H\"uckel  screening length. 

\subsection{Solution for the layered geometry}
\label{sec:tf}

In the layered geometry the differential Eq.~(\ref{eq:ede}) reduces to
a one-dimensional problem and can be readily solved. The geometry is 
identical as in the UDA approximation (Fig.~\ref{fig:dropdr}). The central
$B$ layer has width $2 R_d$ and the cell has width $2 R_c$. The $r$ 
coordinate is perpendicular to the layers and $r=0$ corresponds to the 
center of the $B$ layer. By symmetry the field is zero at $r=0$ and $r=R_c$.
  In this case  the  boundary condition ${\bf E_{\perp}}=0$ 
for the electric field perpendicular to the surface at the cell
boundary automatically warrants the  neutrality condition 
[Eq.~(\ref{eq:neut})] due to Gauss theorem. 

Apart from the cell boundary 
the cell itself has an internal boundary that divides $A$ and $B$ phases
we call $E_0$ the electric field at the $A$-$B$ boundary. The value of $E_0$
is also optimized and this provides an additional boundary condition.

The solution is of the form:
\begin{eqnarray}
E_A(r)&=&E_0 \frac{\sinh[(r-R_c)/l_A]}{\sinh[(R_d-R_c)/l_A]}\nonumber\\
&& \label{eq:elas} \\
E_B(r)&=&E_0 \frac{\sinh(r/l_B)}{\sinh(R_d/l_B)}\nonumber
\end{eqnarray}
where  $E_A(r)\equiv E(r)$  for ${r}\in A$, etc. 

The charge density  is given by:
\begin{eqnarray}
\rho_A&=&\frac{E_0}{4 \pi l_A}
\frac{\cosh[(r-R_c)/l_A]}{\sinh[(R_d-R_c)/l_A]}\nonumber\\
&& \label{eq:nlas} \\
\rho_B&=&\frac{E_0}{4 \pi l_B} \frac{\cosh(r/l_B)}{\sinh(R_d/l_B)}\nonumber
\end{eqnarray}

The electric field at the $A$-$B$
boundary can be related to the jump in the density at the
interface:
\begin{equation}
\label{eq:e0}
E_0=\frac{-4\pi e[n_B(R_d)-n_A(R_d)]}
{[l_B \tanh (R_d/l_B)]^{-1} + \{l_A\tanh [ (R_c-R_d)/l_A]\}^{-1}}
\end{equation}
It plays the same role as the parameter $\delta$ in
Sec.~\ref{sec:freee} so that we can find the optimum charge
distribution between $A$ and $B$ by minimizing the free energy with
respect to $E_0$. 

After replacing  Eqs.~(\ref{eq:nlas}),(\ref{eq:e0}) in the
expression for the free energy [Eq.~(\ref{eq:fdndr})] and minimizing
respect to $E_0$ we find:
\begin{equation}
\label{eq:sole0}
E_0=\frac{4 \pi e \delta_0 \left[ {l_B}^2
\left(n' -1  \right)  - {l_A}^2 n' \right]}
{l_B/\tanh(x R_c/l_B)+l_A/\tanh[(1-x)/l_B]}
\end{equation}
where $\delta_0$ and $n'$ are defined as in Sec.~\ref{sec:parabolic} and
$R_d$ has been eliminated in favor of  the volume fraction 
with $R_d= x R_c$. 

\begin{figure}[tbp]
\epsfxsize=9cm
$$
\epsfbox{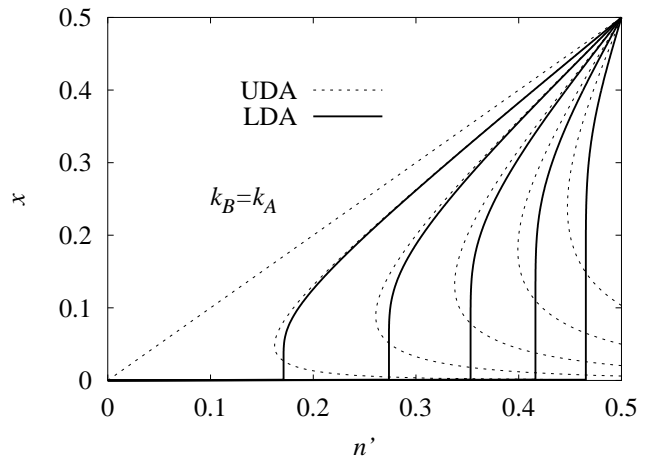}
$$
\caption{ Volume fraction vs. $n'$ for 
(from left to right at the bottom) 
$\lambda=0,0.1,0.2,0.3,0.4,0.5$ and  $k_B=k_A$ in the LDA (thick 
line) and the UDA (thin  line). Only the lower left corner of the plot
is shown since the upper right corner is symmetric by phase exchange.  For the
UDA approximation the lower branch is unphysical 
like in Sec.~\protect\ref{sec:uda}}
\label{fig:xdntf}
\end{figure}

At this point the total free energy per unit
volume $f\equiv F/V$ takes the form:
\FL
\begin{eqnarray}
\label{eq:intf}
f&=&f_A^0+\delta_0 \mu_0 n' + \frac{\sigma}{R_c}\\
&+&2\pi e^2 \delta_0^2 [l_A^2 (n')^2 (1-x) +l_B^2 x (1-n')^2]\nonumber\\
&-&\frac{2 \pi\delta_0^2 e^2 [-l_B^2 (1-n')-l_A^2 n']^2}
{R_c \{l_B/\tanh(x R_c/l_B)+l_A/\tanh[(1-x)R_c/l_A] \} }
\nonumber
\end{eqnarray}
The first two terms are the MC free energy, the third term is the 
surface energy and the last two 
terms are both contributions due to the shift from the MC densities and 
due to the electrostatic energy.

The last step is to minimize  this free energy with respect to the volume 
fraction and
$R_c$. This gives two equations which can be solved numerically 
for $R_c$ and $x$. As in Sec.~\ref{sec:parabolic} it is easier to fix  
$x$ and solve for  $R_c$ and $n'$.

In the following we present results for the case  $k_B=k_A$ and compare 
 with the linearized UDA of 
Sec.~\ref{sec:parabolic} for the layered geometry.

\begin{figure}[tbp]
\epsfxsize=9cm
$$
\epsfbox{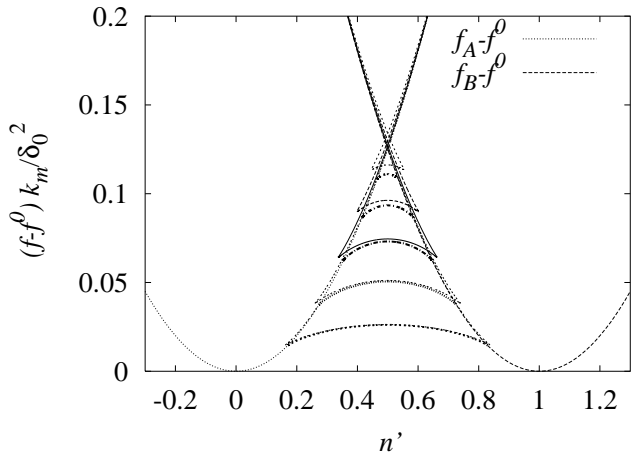}
$$
\caption{
$f_A-f^0$, $f_B-f^0$, and $f-f^0$ in the layered  solution  
for   $\lambda=0.1,0.2,0.3,0.4,0.5$ (from bottom to top) and $k_B=k_A$ 
vs.  $n'$ in the LDA (thick line) and the UDA (thin  line). Here 
$f^0$ is the MC free energy for $\lambda=0$  (a straight line). }
\label{fig:fdntf}
\end{figure}

In Fig.~\ref{fig:xdntf} we plot
the volume fraction as a function of global density in the LDA
approximation and the UDA approximation.
Clearly the result are very similar even quantitatively. In the UDA
there is a jump on the volume fraction form zero to a finite value. 
In the LDA 
the volume fraction is not discontinuous but grows very rapidly at the
threshold for the appearance of the inhomogeneous state. Another 
important difference is that the solutions are not any more
multivalued in the LDA.      

\begin{figure}[tbp]
\epsfxsize=9cm
$$
\epsfbox{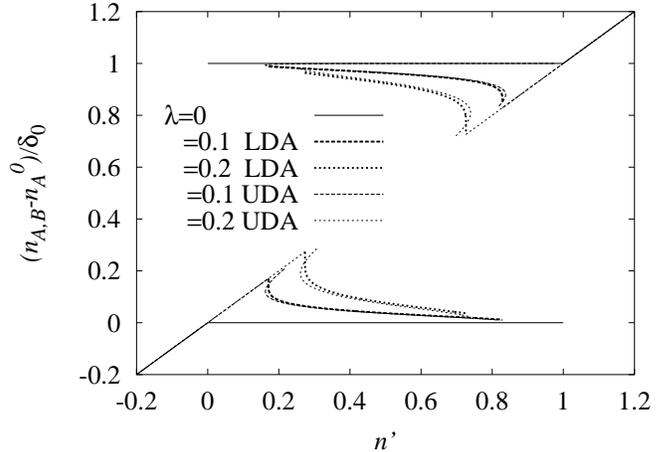}
$$
\caption{
Normalized spatially averaged densities of each phase as a function of
normalized global density $n'$ for different $\lambda$, 
$k_B=k_A$ and the linearized UDA (thin lines) and the LDA (thick lines).
The lower curves  correspond to $A$ phase and the upper curves 
to $B$ phase. In the coexistence region multivalued densities 
appear in the linearized UDA. The long branch is the physical one.  }
\label{fig:nanbdntf}
\end{figure}

In Fig.~\ref{fig:fdntf}  we show $f_A(n')$, $f_B(n')$ and the total
free energy with $k_B=k_A$
for various $\lambda$. The   MC line $f^0(n')=f_A^0+n'(f_B^0-f_A^0)$
has been subtracted. The behavior of the layered solution in the UDA
is similar to the one found for drops in Sec.~\ref{sec:uda} and coincides 
with it at small $\lambda$.  In the
LDA approximation multivaluation disappears. 
The relaxation of the UDA approximation produces 
obviously a gain in energy since the functional that we are minimizing 
is the same in LDA and the UDA but in the UDA we are imposing an extra
constrain on the densities. The gain  in energy however is quite
small. The phase diagram in the
UDA and the LDA (not shown) are both very similar (even
quantitatively) to the one
for drops of Sec.~\ref{sec:uda} except that they are fully symmetric.  
The critical $\lambda$ above which 
the inhomogeneous solution is never stable is given for $k_B=k_A$ by 
$\lambda_c=(9/5)^{1/3}/2\sim 0.61$ in the LDA and by $\lambda_c\sim 0.70$
in the UDA.

In Fig.~\ref{fig:nanbdntf} we show the densities in each phase in the 
UDA. This is compared with the densities of each phase in the LDA 
averaged spatially over the space spanned by each phase. 
Again the behavior is remarkably similar and the density discontinuities
of the UDA 
become  very steep changes with LDA.

\begin{figure}[tbp]
\epsfxsize=9cm
{\hspace{0.cm}{\psfig{figure=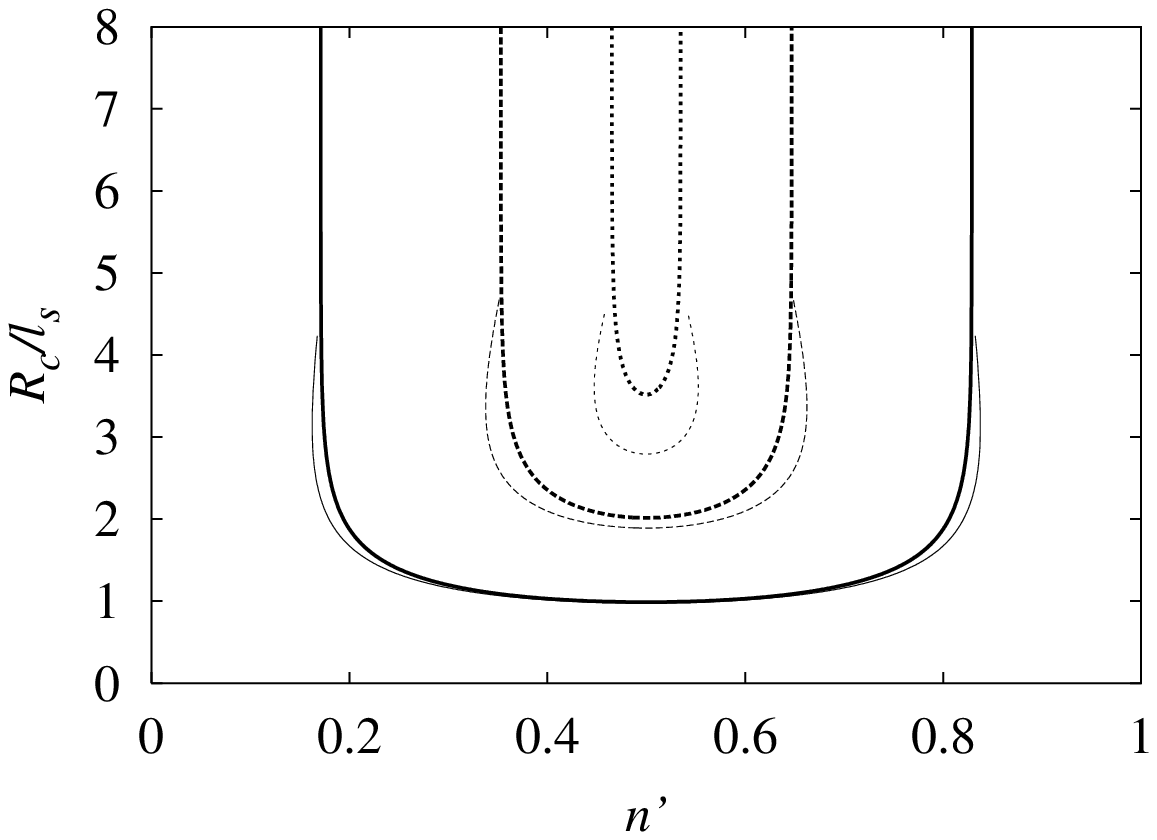,width=9cm}}}
{\hspace{0.cm}{\psfig{figure=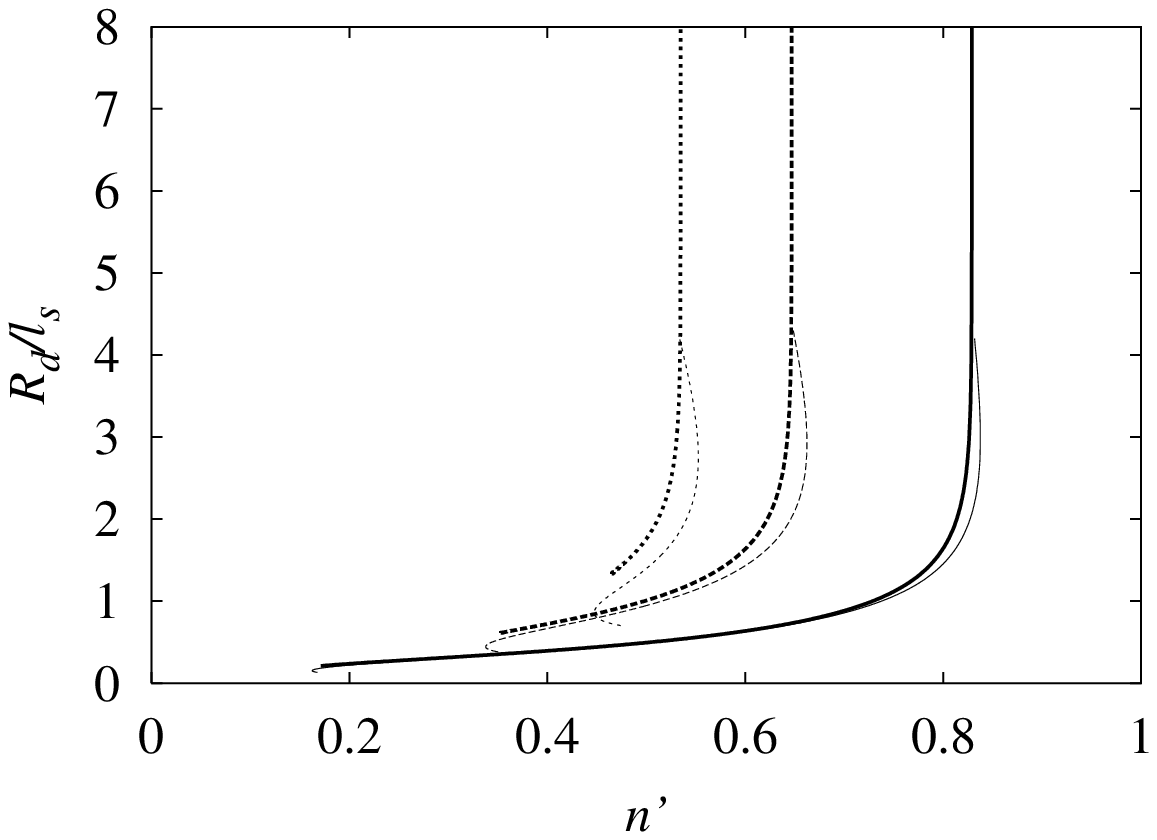,width=9cm}}}
\caption{$R_c$ and  $R_d$ in units of the screening length  $l_s$
defined above  Eq.~(\protect\ref{eq:la}) vs.  $n'$. 
in the linearized UDA (thin lines) and the LDA
(thick lines). We show the 
curves for  $\lambda=0.1,0.3,0.5$ 
which  increases from  bottom to top in the top panel and from 
right to left at the top in the lower panel. 
}
\label{fig:rcdntf}
\end{figure}

Finally in Fig.~\ref{fig:rcdntf} we show the behavior of the
dimensions of the cell and of the $B$ layer as a function of global density. 
Due to perfect phase exchange symmetry the cell width $2R_c$ as a
function of $n'$ is
symmetric and has the minimum exactly at $n'=0.5$. 
The discontinuous jump at the threshold in the UDA becomes a
divergence in the LDA. For the same parameters 
the cell width are  smaller in the UDA than in the LDA. 
This can be understood by noticing that in the UDA  the 
widths are of order $l_s= [\sigma/(\delta_0 e)^2]^{1/3}$.
Roughly speaking we can say that the  effect of the LDA is: i) to increase
the surface energy due to the bending of the charge distributions at
the surface and ii) to screen the electric fields which can be 
schematized 
as an effective reduction of the charge $e$. Both effects tend to 
increase the with of the layers as found.  

For small $\lambda$ Fig.~\ref{fig:rcdntf} shows that the LDA and UDA
radius coincide just as the full solution. 
 This is because $l_d \sim \sqrt{\lambda} l_s<< l_s$
[c.f. Eqs.~(\ref{eq:ld}),(\ref{eq:la})] so that the density is almost 
constant inside
the layer even in the LDA and the solutions are virtually the same.
In this case the Thomas-Fermi approximation is ineffective to
generate a surface energy since all surface energy effects other that the
ones explicitly included in $\sigma$ are due to density variations. 
In other words if one sets $\sigma=0$ the system prefers to make small drops 
to avoid both the Thomas-Fermi induced surface energy effect and the 
Coulomb cost. This however is a drawback of the Thomas-Fermi approximation 
since small drops will certainly  have a large surface energy due to the 
confinement of the electron gas.  It is well known 
that Thomas-Fermi theory is a poor approximation to model 
surfaces.\cite{jon73}

If one increases  $\lambda$  inhomogeneities are possible until the point 
in which $l_d\sim l_s$ and $\lambda=\lambda_c$. It is not possible to have 
inhomogeneities of dimension $l_d>> l_s$ because in the region far
from the surface screening makes the local density to coincide with the 
global density and this inhibits any PS  energy gain. It is then 
convenient for the  system to avoid any surface  and remain single phase.

\begin{figure}[tbp]
\epsfxsize=9cm
$$
\epsfbox{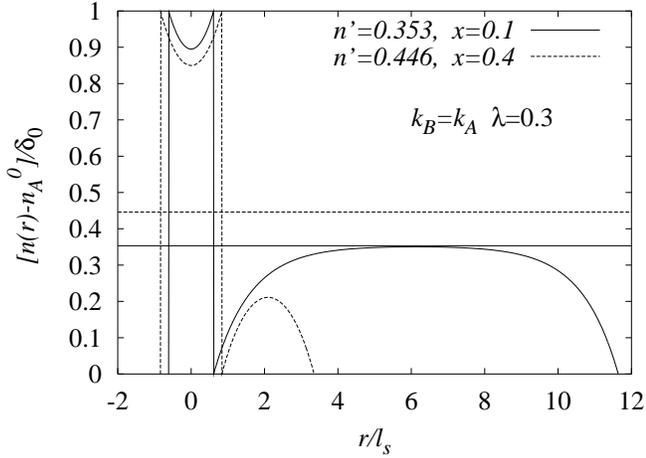}
$$
\caption{Density profile for $\lambda=0.3$ and different values of the
global density. The region close to $r=0$ corresponds to $B$
phase and the rest is $A$ phase. The structure repeats  periodically
in the $r$ direction. The horizontal lines signal the global density.  
}
\label{fig:ndr}
\end{figure}

In Fig.~\ref{fig:ndr} we show the density profile for  $\lambda=0.3$
and for two different values of the global density. One is close to
the threshold  for the appearance of $B$ phase ($n'=0.353$). 
In this case the
$A$ density is close to the density of the background and bends down 
close to the interface to screen the $B$ layer charge. Well in the bulk of
$A$ phase, where the charge density coincides with the density of the
background, we have $E\sim 0$ as expected for a metal.  
When the global density increases the local densities decrease
according to the behavior discussed before for the average densities 
(Fig.~\ref{fig:nanbdntf}). The layers become of the order of the
screening length and the electric field is never completely
screened.

\section{Conclusions}
\label{sec:con}
In this work we have generalized the Maxwell construction to a situation that
appears often in strongly correlated electronic systems, i.e. phase separation 
frustrated by the LRC interaction. 

We discussed i) the stabilization of the uniform phases as the frustrating 
forces are increased, 
ii) the anomalous behavior of the frustrated phase separated mesoscopic 
state and iii) the singular behavior which results in 
a lattice instability when frustration dominates.

 We used a UDA and a more involved LDA approach. Both are shown to give
very similar results thus  justifying in general the use of the much simpler 
UDA.  
For the LDA we have approximated the energy functional in the case of a metal
with the the simplest LDA functional i.e. the Thomas-Fermi approximation.
Our formulation however is general and allows for more sophisticated 
functionals.   

As it is intuitively expected, the LRC interaction tends to stabilize 
the non-separated uniform phases in the presence of a rigid background. 
This has been illustrated in the general analysis 
of two generic phases described by parabolic free energies.
We have shown that  the region of phase separation contracts when the LRC and
surface energy effects are  switched on and disappears above a critical value 
of a dimensionless parameter $\lambda$.
This parameter plays the role of an effective coupling and
characterizes the competition between the energy cost due to the
surface and Culombic energy  and the energy gain in the MC i.e. 
it controls the degree of frustration. 
The balance between these energies 
determine whether the phase separated state exists or not.

When $\lambda$ is small ($\lambda<\lambda_c$)  a mixed state arises.
We have modeled this situation by considering  a Wigner crystal of drops of one
phase hosted by the other phase and a layered geometry which behaves  as one
dimensional analog of the  Wigner crystal. 
We believe that our general 
conclusions (including the existence of a critical $\lambda$) 
 are not sensitive to the geometry of the mixed
state as long as the two length scales $R_c$ and $R_d$ are present and 
both are much larger than the interparticle distance. 
The former length (cell size) characterizes a 
periodic structure and the latter (bubble size) how this periodic structure is 
divided to host the two phases. An indication that the geometry is not 
very important 
comes from the fact that the plots of the physical quantities in 
Sec.~\ref{sec:parabolic}  for $k_A=k_B$ are quite symmetric 
to an exchange of the two phases, each one having a different shape. 
This means that 
the behavior of the drops is not much different from the behavior of their 
counterpart, the interstitial regions.  The same happens when
comparing the behavior of drops and layers.

In the mixed state novel non-linear effects appear which are not 
present in the unfrustrated MC. Within the UDA 
the volume fraction and the drop radius of the minority phase  
do not start from zero but from a finite value and
the transition to the  drops state  is abrupt. In the LDA 
physical quantities are not discontinuous but grow very steep at the
threshold mimicking the discontinuous behavior.

A further non-linear effect in the drop state is that 
the local densities of each phase have an anomalous 
behavior decreasing as the global density increases. 
This can affect properties of the
system which are sensitive to the local density and will be illustrated
in paper II with the  Curie temperature of the manganites. 
We emphasize that also local probes like NMR, core spectroscopy 
etc. should be sensitive to this effect and may be 
used to detect Coulomb frustrated phase separation in real systems.

In the case of strong Coulomb interaction and large surface energy 
($\lambda>\lambda_c$)  a transition between two uniform phases 
 occurs. 
We have shown that in this case the compressibility is singular and a
lattice instability will take place if the ionic background is not
fully rigid. The system (both electrons and ions) can separate in two 
neutral phases with different specific volumes.

In principle also
at the transition point to the drop state a lattice instability 
can arise for the same reasons discussed in the  
$\lambda>\lambda_c$ case, 
although the instability is now much weaker.

When do we expect such a mesoscopic phase separation to prevail against 
microscopic phase separation (like stripes)?. In order to have mesoscopic
phase separation we need that the  
interparticle  distance ($\sim n^{-1/3}$)  be smaller than the
inhomogeneous length $l_d$ which should be smaller than the screening 
length $l_s$. This implies:
$$ n^{-1/3}<\left(\frac{  \sigma\epsilon_0 }{ e^{2} \delta_0^2 }\right)^{1/3}
<\left(\frac{\epsilon_0}{4 \pi e^2 k_m}\right)^{1/2}$$
We see that large dielectric constants favor both large drops and 
inhomogeneous states so polar materials which have typically 
large static dielectric constants ($\epsilon_0 \sim 10 \sim 100$) 
are ideal candidates. Small  $\delta_0$ or large $\sigma$ favors large drops 
but a too small $\delta_0$ or a too large $\sigma$ can inhibit 
phase separation at all. Small values of $\delta_0$ can occurs in manganites
where typically a variety of different ground states  with close densities 
compete with each other (see paper II for a specific example). 
This suggest either large drops or total frustration 
with  lattice instabilities close to the transition from one phase 
to the other.   We mention that these 
lattice instabilities, which also involve volume variations, are
reminiscent of the  macroscopic phase separation observed in some 
manganites.\cite{ueh99}

Finally small compressibilities favor the PS states. This suggest 
that these effects can be important for  bad metals or close to
metal insulator transitions.

We believe that to some extent at least some of the effects found here
can survive also in the microscopic frustrated phase separation. In fact for
example, the corrections to the electrostatic energy due to the
discreetness of the charge which are computed in the Appendix ~\ref{app:disc}  
for the very unfavorable case of a classical Wigner crystal can be
irrelevant for small metallic inhomogeneities  due  to quantum blurring.
In this case however
one should take into account the structure of the underling atomic
potential. Of course if quantum blurring effects are two strong 
one should be concern with the stability of the whole  
superstructure against quantum fluctuations.

\appendix
\section{correction due to discreetness of the charge  }
\label{app:disc}
In order to compute the electrostatic energy in the UDA 
[Eq.~(\ref{eq:eelev})]  we
assume that the charge within one drop is spread uniformly.
Variations of density can arise because of screening effects
as discussed
in Sec.~\ref{sec:tf} and because of intrinsic  charge inhomogeneities 
internal to the particular phase. Here we discuss the latter effect.

Let us now  consider an extreme  limit and assume that both phases A and B 
are two classical Wigner crystals of electrons as a prototypical case in
which the charge is intrinsically non-uniform. What is the correction to the 
Eq.~(\ref{eq:eelev})?.

In the host phase we neglected the interaction between the neutral
A Wigner crystal of electrons and a background of charge density $(n-n_A) e$ 
(see Fig.~\ref{fig:dropdr}).
The fluctuation of the charge inside the crystalline Wigner-Seitz cell 
can make this interaction nonzero. Also for the 
phase forming the drop we have to consider the interaction between 
the neutral B Wigner crystal and  a background of
charge density $(n-n_B) e$.

The electrostatic contribution per drop is: 
$$\epsilon_{A-b}=-\frac{3 e q_A}{10 r_A} N_A$$
with
$$ q_A=\frac{4\pi}3 r_A^3 (n-n_A) e$$ 
and a similar expressions for B phase. Here $\frac{4\pi}3 r_A^3=1/n_A$
and $N_A$ is the number of electrons of A phase in a drop:
$$N_A= n_A v_d (\frac1x-1)$$
$$N_B= n_B v_d $$
The total contribution per unit volume to the electrostatic energy is:
\begin{equation}
\label{eq:eb}
\Delta e_e=\frac{2\pi e^2}5 [(n_B-n) n_B r_B^2  x+  (n_A-n) n_A r_A^2 (1-x) ]
\end{equation}
Clearly [c.f. Eq.~(\ref{eq:eelev})] the correction $\Delta e_e/e_e$
is of order $r^2_{A,B}/R_d^2$ so it is negligible unless
the volume of the drop is of the order of the volume per particle in which 
case the whole computation has no sense.

\section{``Metallic'' drops in ``vacuum''}
\label{app:ka0}

In this appendix we discuss in detail the case of a compressible phase ($B$) 
growing in an incompressible phase ($k_A=0$). 
This simplifies the physics because
the $A$ density is fixed so there is not interchange of particles
 and the $B$ density is not any more bivaluated
for small densities as shown in  the lower panel of Fig.~\ref{fig:nsnldn}.
We present an alternative treatment of the frustrated phase separation 
phenomena which enlightens the underlying physics
 and discuss the pressure exerted by  the mixing forces in detail.

To fix ideas we  call  B phase
a ``metal'' and A phase  the ``vacuum''. Accordingly we
put $n_A=n_A^0=0$ and $f_A=0$. 
These last conditions do not change the solution but make
the interpretation more transparent. 

Since in this case the number of particles in each phase is fixed 
(zero for the vacuum) we can minimize the energy per particle 
$E\equiv f/n$ given by: 
\begin{equation}
\label{eq:epp}
E=\frac{f_B}{n_B}+ \frac{e_m(x,n_B)}{n}  
\end{equation}
This has to be minimized respect to the volume fraction 
taking into account that the density $n_B$ is also a function 
of the volume fraction given by the constrain $n_B=n/x$. 
By putting the  derivative respect to $x$ of Eq.~(\ref{eq:epp}) equal to 
zero we obtain:
\begin{equation}
\label{eq:pl}
p_B= \frac{\partial e_m}{\partial x}- \frac23 \frac{e_m}{ x}
\end{equation}
The left hand side originates in the  first term in Eq.~(\ref{eq:epp}) 
and is the intrinsic pressures of the metal, i.e the 
pressure that the metal exerts on the surface. The right hand side 
is the pressure that the mixing forces, considered as ``external'' to the 
inhomogeneity,  exert on the metal. We call this 
the mixing pressure ($p_m$). In equilibrium both pressures balance 
($p_B=p_m$). 

The mixing pressure has two terms, the first [right hand side of 
Eq.~(\ref{eq:pl})] comes from the explicit dependence of the mixing energy 
on the volume fraction at constant  $n_B$  and is proportional to  $u'(x)$.
For an ideally 
symmetric $u(x)$ (see Fig.~\ref{fig:udx}) this term is positive 
for $x<0$ and is negative for $x>0$. We can say that  this term tends 
to ``compress'' the metal in a less than half filled cell  
whereas it tends to ``stretch'' the metal (negative pressure) in the opposite 
case.  This is just the expected tendency 
of the mixing energy to favor the closest uniform phase ($x=1$ or $x=0$). 
The second term is due to the  dependence of the mixing energy on the volume 
fraction through  $n_B$ at constant particle number. An expansion of 
the cell at constant particle number produces a decrease on $n_B$ 
which reduces the mixing energy [Eq.~(\ref{eq:em})].  This produces a 
negative pressure contribution proportional to $-u(x)/x$. The net contribution
is given by $u'(x)-2u(x)/(3x)$ [Eq.~(\ref{eq:pl})].  It follows that 
for more that half filled cells the metal is subject to a net negative 
pressure and for less than half filled cells the metal is subject to 
negative or positive pressures depending on the geometry and the 
volume fraction.  The upper  curves in 
Fig.~\ref{fig:nsnldx} are proportional to $u'(x)-2u(x)/(3x)-1$. 
For drops the mixing pressure is positive for small $x$ and then 
becomes negative
whereas for layers the mixing pressure is negative for all $x$. 

\begin{figure}[tbp]
\epsfxsize=9cm
$$
\epsfbox{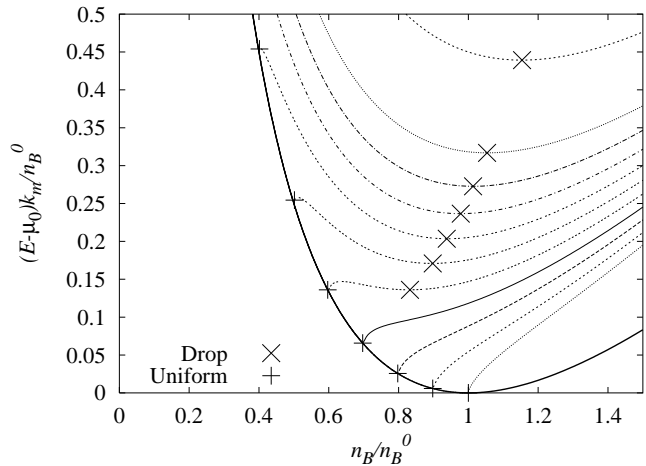}
$$
\caption{Normalized energy per particle as a function of $n_B/n_B^0$
for $\lambda=0.3$.  The thick line correspond the uniform phase
and the thin lines to the drop state with  $n'$ changing from zero 
(top) to one (bottom) in steps of 0.1. The crosses indicate the 
drop solution and the uniform solution. }
\label{fig:fdnb}
\end{figure}

The appearance of negative pressures indicates that the metal  can be stable
at densities  which in the absence of LRC forces would be unstable so it is 
an indication of the stabilization effect of the LRC forces. 
In Fig.~\ref{fig:fdnb} we show $E(n_B)$ for a parabolic free energy.
The intrinsic pressure ($\propto d E/d n_B$) is negative for $n_B<n_B^0$ 
and is positive 
for   $n_B>n_B^0$. We will show below that stable solutions can be found 
in the region $n_B<n_B^0$ which are inaccessible (unstable) according to MC. 

The following example clarifies the physical mining of the negative pressures. 
Consider a neutral liquid 
with short range attractive forces. At negative pressure molecules will be at
distances larger than the equilibrium distance and this implies an energy cost 
proportional to the volume. The system can relax by creating a surface and 
relaxing all molecules to the equilibrium distance. The energy cost 
proportional to the surface is much less than the energy gain proportional to 
the volume and this produce the MC instability. In the presence of mesoscopic 
frustrated PS this can 
not be done because for the drops the surface is not any more 
negligible respect to the volume. In fact the optimum drop ratio can be seen
as the length scale at which successively breaking large drops subject to the 
negative mixing  pressure is not any more convenient due to the surface 
energy cost.

In the following we illustrate the behavior of the solution performing
a graphical minimization of the energy  for the  drop geometry and the 
parabolic free energy Eq.~(\ref{eq:flfs}). 
 Instead of minimizing with respect to the volume fraction 
we use the constraint to eliminate
the volume fraction in favor of $n_B$ ($x=n/n_B$).
The energy per unit particle is given by: 
\FL
\begin{eqnarray}
\label{eq:emmu}
E-\mu^0 &=& \frac{n^B_0}{k_m} \Bigg\{  \frac{(n_B-n_B^0)^2}{2 n_B n_B^0}\\ 
&+&    \frac3{2^{4/3}} \lambda  \left(\frac{n_B^0}{n_B}\right)^{1/3} 
\left[ 2 - 3 {\left( \frac{n}{n_B} \right) }^{1/3} 
+ \frac{n}{n_B} \right]^{1/3} \Bigg\}\nonumber
\end{eqnarray}
The $f_0^B$ term has been eliminated with the MC condition. 
The first term in the curly brackets  is the  bulk energy contribution.  
The mixing energy per particle 
is $e_m/(x n_B)\sim u(x)/(x n_B^{1/3})$ and contributes to the last
term in the curly brackets.  The geometric factor   $u(x)/x$ gives the 
term in the square brackets.  

The equilibrium density is found by minimizing 
Eq.~(\ref{eq:emmu})
with respect to $n_B$. In Fig.~\ref{fig:fdnb} we show $E-\mu^0$ as a function
of $n_B$ for $\lambda=0.3$ and different values of $n'$. The thick line is 
the energy of the uniform metal [the first term in the brackets in 
Eq.~(\ref{eq:emmu})] and is minimized at the MC density $n_0^B$.

For very small $n$ (or $x$) the geometric factor is constant and the 
mixing energy contribution  goes as $1/{n_B}^{1/3}$. 
This shifts the minimum to values of the 
density larger than $n_0^B$ as can be seen from the 
upper curves of  Fig.~\ref{fig:fdnb} where the energy per particle
is given for various values of the global density and $\lambda=0.3$. 
This is due to the positive pressure exerted by the mixing energy
of drops  at small volume fraction and  explain the behavior of the 
$n_B$ density in the limit $n'\rightarrow 0$ (Fig.~\ref{fig:nsnldn}).  
As the density increases the density dependence of the  geometric factor
tends to reduce the minimum to lower densities.

As a by product this computation illustrates the  stabilization of a
uniform solution by the long range interaction and the first order
like nature of the transition. 
Above $n'\sim 0.6$ the uniform solution becomes suddenly  more favorable
(see also Fig.~\ref{fig:nsnldn}). Notice that this density is well
inside the MC coexistence region   ($0<n'<1$) showing the
uniform solution stabilization effect.

It is important to remark that the whole behavior can change if
the surface energy $\sigma$ had a strong density dependence.
For this reason the interpretation of $B$ phase as a metal
should be taken with caution since in general in a metal the surface
energy will depend strongly on density. Specific examples will be treated
in paper II.


\begin{references}
\bibitem{notequa} Of course quantum effects can not be neglected at the microscopic 
level of the PS phenomena  but can be neglected at the mesoscopic
level of the large clusters of particles forming the inhomogeineties.
\bibitem{note} The usual definition of compressibility is given by
$\kappa=-V^{-1} d V/ dp$. This is related to our compressibilities 
by $k=n^2 \kappa$.
\end{references}

\end{document}